\documentclass[prd,twocolumn,eqsecnum,longbibliography]{revtex4-1}
\usepackage{amsmath,latexsym,amssymb,mathrsfs,amsthm}
\usepackage[colorlinks=true, urlcolor=blue, citecolor=blue, linkcolor=blue, pagecolor=blue]{hyperref}
\usepackage{graphicx}

\begin{document}
\title{Singularities in Spherically Symmetric Solutions with Limited Curvature Invariants}
\author{Daisuke Yoshida}
\email{d.yoshida@physics.mcgill.ca}
\affiliation{Department of Physics, McGill University, Montr\'eal, Qu\'{e}bec, H3A 2T8, Canada}

\author{Robert H. Brandenberger}
\email{rhb@hep.physics.mcgill.ca}
\affiliation{Department of Physics, McGill University, Montr\'eal, Qu\'{e}bec, H3A 2T8, Canada}

\begin{abstract}
We investigate static, spherically symmetric solutions in gravitational theories which have
limited curvature invariants, aiming to remove the singularity in the Schwarzschild space-time. 
We find that if we only limit the Gauss-Bonnet term and the Ricci scalar, then the
singularity at the origin persists. Moreover we find that  the event horizon can develop a 
curvature singularity. We also investigate a new class of theories in which all components of the 
Riemann tensor are bounded. We find that the divergence of the quadratic curvature
invariants at the event horizon is avoidable in this theory. However, other kinds of singularities 
due to the dynamics of additional degrees of freedom cannot be removed, and the space-time
remains singular.
\end{abstract}

\maketitle
\section{Introduction}

The space-time singularity is one of the most important signs that Einstein gravity has to be
modified at high energies. The singularity theorems~\cite{Penrose:1964wq,Hawking:1967ju,Hawking:1973uf} state that space-time singularities are inevitable in Einstein gravity provided that gravity
is coupled to matter which obeys energy conditions which are natural from the point of view
of classical physics (there are some additional technical assumptions which are
automatically satisfied in the symmetric space-times we are considering). There are many
arguments supporting the view that the Einstein action can only be a low energy effective theory
for gravity. First, it is not a renormalizable theory, and hence cannot yield a consistent quantum
theory in the ultraviolet. Gravitational interactions will inevitably lead to higher curvature
correction terms to the action. Similarly, gravitational interactions of matter field will lead
to correction terms in the effective action for gravity. It is a long-standing hope that curvature
singularities will be removed in a consistent quantum theory of gravity. Specifically, one
could hope that the two most famous gravitational singularities, the Big Bang singularity
of homogeneous and isotropic cosmology, and the Schwarzschild singularity at the center
of a spherically symmetric black hole metric, will be removed in a complete theory of
quantum gravity.  

In this paper, we will explore the question of singularity removal at the level of modified
effective gravitational actions. If we were able to construct a gravitational theory without 
singularities, it would provide a candidate for an effective theory of a consistent theory
of quantum gravity. 

In the context of cosmology, various scenarios to obtain a nonsingular Universe have been investigated.
Inflation was initially proposed as a candidate for a non-singular cosmology \cite{Starobinsky:1980te}. 
The simplest way to obtain an inflationary cosmology is to maintain the Einstein gravitational
action and to assume the presence of a scalar field whose potential energy can lead
to almost exponential expansion \cite{Guth:1980zm}. However, it was shown that such a
scalar field-driven inflationary universe has an initial time 
singularity~\cite{Borde:1993xh,Borde:1996pt} if the scalar field matter satisfies the null energy condition 
It was also shown that an inflationary Universe which is described by the usual spatially flat 
Friedmann- Lema\^{i}tre - Robertson-Walker (FLRW) coordinates is past 
incomplete~\cite{Borde:2001nh}, and hence singularity freeness cannot be discussed 
restricting attention to this coordinate region. 

Nonsingular cosmological background space-times (which might even lead to alternatives to 
inflation as a theory of cosmological structure formation, e.g. the ``matter bounce''
scenario \cite{Finelli:2001sr}) have been constructed in the context of Einstein gravity by
invoking matter which violates the null energy condition. There are models
with a cosmological bounce (see e.g. ~\cite{Novello:2008ra,Brandenberger:2012um,Brandenberger:2012zb,Cai:2014bea,Battefeld:2014uga,Brandenberger:2016vhg} for reviews on bouncing cosmology ), or ``genesis'' models such as Galileon Genesis ~\cite{Creminelli:2010ba,Creminelli:2012my,Nishi:2014bsa,Pirtskhalava:2014esa,Nishi:2015pta,Kobayashi:2015gga,Nishi:2016wty,Libanov:2016kfc} . However, a generic instability for non-singular bouncing solutions was proven in Refs.~\cite{Kobayashi:2016xpl,Akama:2017jsa} in a class of scalar-tensor theories, the so called Horndeski theories~\cite{Horndeski:1974wa,Deffayet:2011gz,Kobayashi:2011nu} and its multi-field extensions~\cite{Padilla:2012dx}. Stable non-singular solutions have then been investigated 
in the framework of scalar-tensor theory~\cite{Ijjas:2016tpn} which goes beyond the framework
in which the assumptions of the no-go theorems have been derived, and it also goes
beyond the usual effective field theory approach
to gravity~\cite{Cai:2016thi,Creminelli:2016zwa,Cai:2017tku}.  

Another way to obtain a non-singular cosmology is to consider higher curvature corrections
\cite{Stelle:1976gc,Stelle:1977ry} as in the Starobinsky model~\cite{Starobinsky:1980te} of 
inflation.  An example of such a higher derivative gravity model aiming to remove ther singularity is 
the infinite derivative gravity model of \cite{Biswas:2005qr,Biswas:2011ar,Conroy:2016sac}, where 
the theory includes all powers of derivatives of the Ricci scalar. In this paper we would 
like to focus on another possibility of obtaining a non-singular gravitational theory with 
higher curvature terms which was proposed by Refs.~\cite{Mukhanov:1991zn, Brandenberger:1993ef}, 
and called the ``limiting curvature construction''. It is a gravitational theory in which extra
terms are added to the Einstein action with the purpose of limiting certain curvature scalars. 
The idea of the construction is to limit one scalar curvature polynomial to finite values 
by introducing a Lagrange multiplier scalar field and adjusting its potential. In this way, we can 
limit any number of curvature scalars by introducing the corresponding number of Lagrange
multiplier scalar fields. However, the difficulty comes from the fact that there are an 
infinite number of curvature polynomials. Thus even if we ensure that a finite number of 
curvature polynomials, e.g. $R$ and $R_{\mu\nu}R^{\mu\nu}$, have finite values, other 
curvature polynomials, e.g. $R_{\mu\nu\rho\sigma}R^{\mu\nu\rho\sigma}$, could possibly diverge. 
Thus, the choice of which curvature polynomials to bound is very important. 

In the case of homogeneous and isotropic space-times, then since the Riemann tensor 
is given by the Hubble function $H$ and its derivative $\dot{H}$, the finiteness of the 
Riemann tensor is ensured if we control these two quantities. However, this is not
sufficient to remove all singularities. It is possible to have geodesically incomplete
space-times where no curvature invariant blows up. The idea in 
\cite{Mukhanov:1991zn,Brandenberger:1993ef} was to adjust the Lagrange multiplier
construction such that at high curvature the cosmological solutions approached a known non-singular
solution, namely de Sitter. Non-singular cosmological solutions based on the limiting curvature 
construction have been investigated in 
Refs.~\cite{Mukhanov:1991zn,Brandenberger:1993ef,Yoshida:2017swb}. The background 
dynamics of a contracting Universe was first studied in Refs.~\cite{Mukhanov:1991zn,Brandenberger:1993ef} and then that of an expanding Universe corresponding to inflationary and genesis scenarios 
was studied in Ref.~\cite{Yoshida:2017swb}. It was also shown that cosmological solutions 
are stable in a wide region of cosmological history.

If the limiting curvature theories are to give a good guide to the ultimate quantum theory
of gravity, they should not only work well for cosmological situations, but also be able
to remove other kinds of singularities appearing in Einstein gravity such as the Schwarzschild 
singularity.  The first example of a non-singular black hole space-time was given by Bardeen 
as a solution of the Einstein-Maxwell theory (see Ref. \cite{Ansoldi:2008jw} for a review of 
Bardeen's model and other non-singular black hole solutions). Motivated by the recent 
developments in modified theories of gravity, non-singular spherically symmetric solutions have 
been investigated also in the context of modified gravity, for example in $F(R)$ gravity with 
an anisotropic fluid~\cite{Olmo:2015axa} and in mimetic gravity~\cite{Chamseddine:2016ktu, BenAchour:2017ivq}. 
Since the limiting curvature construction prevents the divergence of curvature invariants, 
it is natural to expect that spherically symmetric solutions of these theories might be non-singular. 
In fact, a non-singular black hole solution in the 1+1 dimensional space-time in the limiting 
curvature theory was obtained in Ref.\cite{Trodden:1993dm}. However, it was never clarified
whether in this construction the Schwarzschild singularity can be removed in 1+3 dimensional 
space-time.  The purpose of this paper is to study whether the 1+3 dimensional
Schwarzschild singularity can be removed in a theory with limiting curvature invariants.
We will hence investigate static, spherically symmetric solutions with various choices of 
controlled curvatures and potentials of the scalar Lagrange multiplier fields.

Our paper is organized as follows. In the next section, we will review the limiting curvature 
construction of \cite{Mukhanov:1991zn,Brandenberger:1993ef} and propose another class of 
theories where each component of the Riemann tensor is controlled. In Section \ref{sec:GB}, 
we will investigate static, spherically symmetric solutions in a theory with bounded 
Gauss-Bonnet term, which is a ghost free subclass of the limited curvature theories. 
We will find that two kinds of singularities remain, one is the Schwarzschild singularity 
and the other is dubbed as a thunderbolt singularity. The appearance of the 
Schwarzschild singularity can be understood since the construction does not bound all 
curvature polynomials. Next, we investigate a theory in which both the Ricci scalar and the 
Gauss-Bonnet term are limited (Section \ref{sec:GBR}). However we will find that the 
Schwarzschild singularity still cannot be removed. In Section \ref{sec:limRiem}, we 
then investigate a theory in which all Riemann curvature tensor elements are
bounded. Then we will succeed to remove the divergence of the quadratic curvature scalars. 
However we will find other kinds of singularities due to the additional degrees of freedom 
generated by higher derivative interactions. The final section contains a summary of our
results and discussions on the difficulty of obtaining non-singular spherically symmetric 
solutions using the limiting curvature construction.

\section{Gravitational Theory with Limiting Curvatures}
\label{ReviewofLC}

Let us review the gravitational theory with limiting curvature scalars proposed in
Refs.~\cite{Mukhanov:1991zn,Brandenberger:1993ef}. The action of this theory is 
given by 
\begin{align}
S \, = \, \frac{M_{\text{pl}}^2}{2} \int d^4 x \sqrt{-g} {\cal L}
\end{align}
with the Lagrangian density
\begin{align}
{\cal L} = R + M_L^2 \left(\sum_{i = 1}^{n}  \chi_{i} I_{i} - V(\chi_{i}) \right), \label{Slimcs}
\end{align}
where $M_{\text{pl}}$ is the reduced Planck mass, $g_{\mu\nu}$ is the space-time metric,  
$R$ is the Ricci scalar of the space-time and $I_{i}$ are dimensionless scalar curvature 
polynomials constructed from Riemann tensor $R^{\mu}{}_{\nu\rho\sigma}$ and their 
covariant derivatives,
\begin{align}
I_i = I_i (g^{\mu\nu}, M_{L}^{-2}R^{\mu}{}_{\nu\rho\sigma}, M_{L}^{-1} \nabla_{\mu} ).
\end{align}
Here, we introduced only a single dimension-full parameter $M_L$ just for simplicity. Note 
that it is natural to expect $M_L = {\cal O}(M_{\text{pl}})$ if we regard the origin of 
the modification terms in the action as a quantum effect of gravity. This theory includes 
$n$ dimensionless Lagrange multiplier scalar fields $\chi_{i}$ and their potential term 
$V(\chi_{i})$, which play an important role in limiting the curvature scalars $I_{i}$. 

From the variations with respect to $\chi_{i}$ we obtain the equations,
\begin{align}
 I_{i} = V_{,\chi_{i}} \, .\label{Ii=Vchii}
\end{align}
If we use a potential whose derivatives are finite for all field values of $\chi_{i}$, only 
solutions with finite curvature scalars $I_{i}$ are consistent with the equations of motion. 
Thus we can eliminate any curvature singularity where one of the curvature scalars $I_{i}$ diverges. 
However since there are an infinite number of curvature scalars constructed from 
$R^{\mu}{}_{\nu\rho\sigma}$ and their derivatives, it is still nontrivial whether curvature 
scalars other than $I_{i}$ are finite or not. For example, if we consider a theory with $n=1$ 
and $I_1 = R$, then the Schwarzschild singularity 
would remain because the Ricci scalar vanishes for Schwarzschild.  

A guideline for the choice of the bounded curvature scalars $I_i$ 
was proposed in \cite{Mukhanov:1991zn,Brandenberger:1993ef}
and called the {\it limiting curvature hypothesis}. The idea is to find some
invariant $I$ which has the property that $I = 0$ has 
only a definite class of non-singular space-times (e.g. de Sitter space-times) as
solutions, and then to choose the potential for the Lagrange multiplier field
associated with $I$ such that at high curvatures $I$ is driven to zero. More
generally, the idea was to force the solution to approach a well-defined nonsingular
space-time when all curvature invariants $I_i$ take their limiting values corresponding 
to $\chi_i \rightarrow \infty$. For example, in the case of homogeneous and isotropic 
FLRW space-time, the Riemann tensor is given by the Hubble function $H$ and its time 
derivative $\dot{H}$. Thus the assumption of the limiting curvature hypothesis is satisfied 
if we control two curvature scalars 
$ I_{1}|_{\text{FLRW}} \propto  H, I_{2}|_{\text{FLRW}} \propto \dot{H} $ by a 
potential that satisfies $V_{,\chi_1} \rightarrow \text{const}$ and $V_{,\chi_2} \rightarrow 0$. 
As investigated in Ref.~\cite{Yoshida:2017swb}, such curvature scalars are realized in 
terms of $R_{\mu\nu}$ and its covariant derivatives. However, this choice of curvature
invariants does not work for vacuum solutions like Schwarzschild 
because $R_{\mu\nu}$ vanishes in the Schwarzschild space-time. 
Thus for our purpose, which is to remove a curvature singularity in a spherically symmetric space-time, we need to consider other curvature scalars that prevent the divergence of $R_{\mu\nu\rho\sigma}R^{\mu\nu\rho\sigma}$. We will investigate  this kind of theories in the sections \ref{sec:GB} and \ref{sec:GBR}. Note that as soon as we abandon the assumption of
homogeneity and isotropy, the dynamical system becomes much more complicated
since the equations are now true partial differential equations. Hence, we should expect
that it is more difficult to prevent singularities.

It should be noted that if the equations \eqref{Ii=Vchii} can be solved for $\chi_i$,
\begin{align}
 \chi_i = \chi_i(I_1, I_2, \cdots, I_n),
\end{align}
one can eliminate $\chi_i$ from our action just by plugging in these solutions. 
Then we obtain pure metric theory including higher derivatives;
\begin{align}
 {\cal L} = R + M_L^2 F(I_1,I_2, \cdots, I_n),\label{F(I)}
\end{align}
where $F$ is given as the Legendre transformation of $V$;
\begin{align}
 F(I_j) = \sum_{i} \chi_{i}(I_j) I_{i} - V(\chi_{j}(I_k)).
\end{align}
Thus the theory with limited curvature can be regarded as a higher curvature 
modification of Einstein gravity. For example, the limiting curvature theory with 
$n=1$ and $I_1 = R/M_{L}^2$ corresponds to $F(R)$ gravity. 
 
Before closing this section, let us suggest a way to limit the curvature without 
assuming any particular symmetry of space-time. This would be complicated
to achieve in the framework of the theory \eqref{Slimcs}, but it easily realized 
if we consider a slightly modified theory 
\begin{align}
{\cal L} = R + \chi_{\mu\nu\rho\sigma} R^{\mu\nu\rho\sigma} - M_{L}^2 V(g^{\mu\nu},\chi_{\mu\nu\rho\sigma}),\label{LCT}
\end{align}
which can be called a gravitational theory with limited curvature {\it tensor}.  
Here a tensor field $\chi_{\mu\nu\rho\sigma}$ is introduced instead of scalar fields 
$\chi_i$.  $V(g^{\mu\nu}, \chi_{\mu\nu\rho\sigma})$ is a scalar function of 
$\chi_{\mu\nu\rho\sigma}$, which controls the Riemann tensor. Variation with respect to $\chi_{\mu\nu\rho\sigma}$ gives the equations,
\begin{align}
 \frac{1}{M_L^2}R^{\mu\nu\rho\sigma} = \frac{\partial V}{\partial \chi_{\mu\nu\rho\sigma}}.\label{Riem=delV}
 \, .
\end{align}
Then we assume 
\begin{align}
 \frac{\partial V}{\partial \chi_{\mu\nu\rho\sigma}} \rightarrow \kappa  g^{\mu[\rho}g^{\sigma]\nu},\label{delVdelchimnrs} 
\end{align}
with a constant $\kappa$ at the limiting values $\chi_{\mu\nu\rho\sigma} \rightarrow \infty$.
Since the right hand side of \eqref{delVdelchimnrs} is nothing but the Riemann curvature of the 
constant curvature space, which is (Anti) de Sitter space-time for positive (negative) $\kappa$ 
or Minkowski space-time for $\kappa = 0$, we know that the solution will approach 
a non-singular space-time at limiting values of the Lagrange multiplier fields  - a conclusion
which holds without assuming any special symmetry. However, it is not clear that the
asymptotic region can be reached without encountering singularities, singularities
which would be different from curvature singularities. We will investigate the spherically 
symmetric solutions of this kind of theory in Section \ref{sec:limRiem}. 

Let us introduce trace and traceless parts of $\chi_{\mu\nu\rho\sigma}$ by,
\begin{align}
 &\chi_{\mu\nu} := \chi_{\mu}{}^{\rho}{}_{\nu\rho},\qquad \chi:= \chi^{\mu}{}_{\mu}, \notag\\
& \hat{\chi}_{\mu\nu\rho\sigma} := \chi_{\mu\nu\rho\sigma} -  (g_{\mu[\rho}\chi_{\sigma]\nu} -g_{\nu[\rho}\chi_{\sigma]\mu}) + \frac{1}{3} \chi g_{\mu[\rho}g_{\sigma]\nu},\label{chi4=}
 \end{align}
 similar to the definition of the Ricci tensor, the Ricci scalar and the Weyl tensor. Then by 
 introducing the traceless part of $R_{\mu\nu}$ and $\chi_{\mu\nu}$ by
 \begin{align}
  \hat{R}_{\mu\nu} &:= R_{\mu\nu} - \frac{1}{4} R g_{\mu\nu},\label{delRhat}\\
  \hat{\chi}_{\mu\nu} &:= \chi_{\mu\nu} - \frac{1}{4} \chi g_{\mu\nu},
 \end{align}
 our action can be written in following form:
\begin{align}
 {\cal L} =&  R + \frac{1}{6} \chi R  + 2 \hat{\chi}_{\mu\nu} \hat{R}^{\mu\nu} + \hat{\chi}_{\mu\nu\rho\sigma}C^{\mu\nu\rho\sigma} \notag\\
& -  M_L^2 V(g^{\mu\nu},\chi, \hat{\chi}_{\mu\nu}, \hat{\chi}_{\mu\nu\rho\sigma} ) \, .
\end{align}
Thus, the variations with respect to $\hat{\chi}_{\mu\nu\rho\sigma}, \hat{\chi}_{\mu\nu}$ and 
$\chi$ give the following equations limiting  the curvature tensors,
\begin{align}
 &\frac{1}{M_L^2}C^{\mu\nu\rho\sigma} = \frac{\partial V}{\partial \hat{\chi}_{\mu\nu\rho\sigma}},\\
&\frac{1}{M_L^2}\hat{R}^{\mu\nu} = \frac{1}{2}  \frac{\partial V}{\partial \hat{\chi}_{\mu\nu}},\\
& \frac{1}{M_L^2}R  =  6 \frac{\partial V}{\partial \chi}.\label{Rsc=Vchi}
\end{align}

Similar to theories with limited curvature scalars, we can write this theory in the form of 
a pure metric theory. In this case, we obtain so-called $F(\text{Riemann})$ 
gravity \cite{Deruelle:2009zk};
\begin{align}
 {\cal L} = R + F(g_{\mu\nu}, R^{\mu\nu\rho\sigma}),\label{FRiem}
\end{align}
where $F$ is a scalar constructed from $g_{\mu\nu}$ and the Riemann tensor 
$R^{\mu\nu\rho\sigma}$, which is related with $V$ via the Legendre transformation,
\begin{align}
 F(R^{\mu\nu\rho\sigma}) = \chi_{\mu\nu\rho\sigma}(R) R^{\mu\nu\rho\sigma} - V(\chi_{\mu\nu\rho\sigma}(R)),
\end{align}
where $\chi_{\mu\nu\rho\sigma}(R)$ is defined as a solution of \eqref{Riem=delV}. Note that the 
equivalence between \eqref{LCT} and \eqref{FRiem} holds only when the 
equation \eqref{Riem=delV} can be solved by $\chi_{\mu\nu\rho\sigma}$.

\section{Spherically symmetric solution with limiting Gauss-Bonnet term}
\label{sec:GB}

\subsection{Ghost Free Higher Derivative Gravity with Riemann Square Invariants}

As we have seen in the previous section, a theory with limited curvature scalars can 
be written in the form of a higher derivative gravitational theory \eqref{F(I)}. In general, 
higher derivative gravity models have pathological ghost degrees of freedom \cite{Stelle:1977ry}. 
The presence of ghosts in higher derivative theories can be shown exactly in the case of 
un-constrained systems. This is known as Ostrogradsky's theorem
\cite{Ostrogradsky:1850fid}. However, there 
is room to construct ghost free higher derivative theory in constrained or gauge systems 
as in gravitation. The simplest example of a ghost-free theory is $F(R)$ 
gravity \cite{Sotiriou:2008rp, DeFelice:2010aj}. Then it was shown that ghost-free higher 
derivative theories can be constructed even if the covariant derivative of $R$ is 
included \cite{Naruko:2015zze}. However these theories are not suitable for the 
purpose of eliminating the Schwarzschild singularity because they allow us to
limit only the the Ricci scalar $R$ and its derivatives and cannot limit $R_{\mu\nu\rho\sigma}R^{\mu\nu\rho\sigma}$, which blows up near the Schwarzschild 
singularity. Thus we need to consider a higher curvature theory which includes at least 
the Riemann square invariant. Note that a non-singular spherically symmetric 
solution is obtained in the framework of $F(R)$ gravity in the presence of an anisotropic 
fluid \cite{Olmo:2015axa}. We will not focus on such a case simply because the 
mechanism to avoid the singularity has nothing to do with our limiting curvature mechanism 
as discussed above.

An example of a ghost free higher derivative gravity with Riemann square term is 
proposed in the appendix of Ref. \cite{Kobayashi:2011nu}. There, it was shown 
that $F(\text{Gauss-Bonnet})$ term is equivalent to a subclass of ghost free 
scalar-tensor theories called Horndeski theories \cite{Horndeski:1974wa}. 
Let us consider Einstein gravity with a $F(\text{Gauss-Bonnet})$ term,
\begin{align}
{\cal L} = R + M_L^2 F({\cal G}/M_L^4),\label{fG}
\end{align}
where the Gauss-Bonnet term ${\cal G}$ is given by,
\begin{align}
 {\cal G} = R^2 -4 R_{\mu\nu}R^{\mu\nu} + R_{\mu\nu\rho\sigma}R^{\mu\nu\rho\sigma}.
\end{align}
By comparing the action \eqref{fG} with \eqref{F(I)}, we conclude that this is a theory 
with limiting curvature scalar $I_1$ with $n=1$ and $I_1 = {\cal G}/M_L^4$. Therefore 
this theory can be written in the form of original limiting curvature theories,
\begin{align}
 {\cal L} = R + M_L^2 \left( \chi \frac{{\cal G}}{M_L^4} -  V(\chi) \right) .\label{SlimG}
\end{align}
Now the Gauss-Bonnet term is controlled by the potential $V$ through the 
variational equation with respect to $\chi$ 
\begin{align}
 \frac{{\cal G}}{M_L^4} = V_{,\chi}(\chi).\label{eomchiGB}
\end{align}
Since the Gauss-Bonnet term includes the Riemann square term which diverges 
at the Schwarzschild singularity, one may hope that the curvature singularity could
be relaxed by forcing ${\cal G}$ to be finite.

\subsection{Spherically symmetric, static, asymptotically flat solutions}
\label{Sec:SSAF}

Let us consider static spherically symmetric solutions of this theory \eqref{SlimG}. The
dynamical variables are the metric tensor $g_{\mu\nu}$ and a single Lagrange multiplier 
field $\chi$. Given the assumption of spherically symmetry, $g_{\mu\nu}$ and $\chi$ 
can be written as
\begin{align}
& g_{\mu\nu}dx^{\mu} dx^{\nu} = - f(r) dt^2 + h(r) dr^2 + r^2 d \Omega^2,\label{sphericallysymg}\\
& \chi = \chi(r),
\end{align}
where $d \Omega^2$ is the metric on the sphere,
\begin{align}
d \Omega^2 = \Omega_{IJ}dx^{I}dx^{J} =  d \theta^2 + \sin^2 \theta d \phi^2 .
\end{align}
Then the Ricci scalar and  the Gauss-Bonnet term can be written as,
\begin{align}
& R(r) = \frac{1}{r^2}\left(1- \frac{1}{h}\right) - \frac{2 f'}{r f h} + \frac{2 h'}{r h^2}+ \frac{f'^2}{2 f^2 h}  + \frac{f' h'}{2 f h^2} - \frac{f''}{f h},\label{R(r)} \\
& {\cal G}(r) = \frac{1}{r^2 \sqrt{f h}}\partial_r \left[ - 4 \sqrt{f h} \frac{f'}{h f}\left(1 - \frac{1}{h}\right)  \right],\label{GB(r)}
\end{align}
where $'$ represents the derivative with respect to $r$.
Making use of these expression, we can write down the action in terms of 
$f, h$ and $\chi$.Then, taking the variation with respect to $f$ and $h$, we obtain the
following equations of motion, 
\begin{align}
&  1 - \frac{1}{h} + \frac{r h'}{ h^2} 
+   2 \frac{h'}{h^2}\left(1 -  \frac{3}{h}\right) \frac{\chi'}{M_L^2} \notag\\ & \qquad 
- \frac{4}{h} \left(1 - \frac{1}{h}\right) \frac{\chi''}{M_L^2} 
- \frac{r^2 M_L^2 }{2}V  = 0,\label{eomfGB}\\
& 1 - \frac{1}{h} - \frac{r f'}{h f} \notag\\
& \qquad -\frac{2 f'}{f h}\left(1 - \frac{3}{h} \right) \frac{\chi'}{M_L^2}
-\frac{r^2 M_L^2 }{2}V =0.\label{eomhGB}
\end{align}
The final equation of motion results from varying with respect to $\chi$ and is 
given by \eqref{eomchiGB}, with the Gauss-Bonnet term given by Eq.~\eqref{GB(r)}. 

In order to limit the Gauss-Bonnet term, we need to use a potential $V$ whose 
$\chi$ derivative $V_{,\chi}$ is finite. As an example of such a potential, here we shall 
focus on the potential
\begin{align}
 V(\chi) = \frac{1}{2} \frac{\chi^2 + 2 \chi^{3}}{1 + \chi^2}.\label{VGB}
\end{align}
The first derivative of $V$ is given by
\begin{align}
 V_{,\chi}(\chi) = \frac{\chi(1 + 3 \chi + \chi^3) }{2 (1 + \chi^2)^2},
\end{align}
which is finite for any $\chi$. Hence, the Gauss-Bonnet term ${\cal G}$ is 
finite through Eq.~\eqref{eomchiGB}. 

First, let us focus on the region $\chi \ll 1$.  There our potential \eqref{VGB} can be expanded as
\begin{align}
 V(\chi) = \frac{1}{2} \chi^2 + {\cal O}(\chi^3).
\end{align}
Then the equation of motion \eqref{eomchiGB} gives the relation
\begin{align}
 \chi = \frac{\cal G}{M_L^4} + {\cal O} \left(\left( \frac{\cal G}{M_L^4}\right)^2 \right). 
\end{align}
Thus, the condition $\chi \ll 1$ corresponds to ${\cal G} \ll M_L^4$. In this region the 
correction terms compared to Einstein gravity can be omitted and then the Schwarzschild 
space-time is a solution.  For the Schwarzschild space-time with mass $M$, the 
Gauss-Bonnet term can be evaluated as
\begin{align}
 \frac{{\cal G}}{M_L^4} = \frac{48 G^2 M^2}{M_L^4 r^6} 
= \left(\frac{r_L}{r}\right)^6,
\end{align}
where $r_L$ is given by
\begin{align}
r_L = 3^{1/6} \left(\frac{4 G M}{M_L^2}\right)^{1/3},
\end{align}
and $G$ is the gravitational constant given by $G^{-1} = 8 \pi M_{\text{pl}}^2$.
Thus the condition $\chi \ll 1$ is equivalent to $ r_L \ll r$.
Here the ratio of $r_L$ to the Schwarzschild radius $r_g = 2GM$ is given by 
\begin{align}
 \frac{r_L}{r_g} &= 
 1.51 * \left(\frac{(2 G M)^{-1}}{M_L} \right)^{2/3}\\
&= 1.38*10^{-25} \left(\frac{M_{\text{pl}}}{M_L}\right)^{2/3} \left(\frac{M_{\odot}}{M}\right)^{2/3},
\end{align}
where $M_{\odot}$ is the solar mass. Thus $r_L \ll r_g$ for the realistic situation, 
$M_L \sim {\cal O}(M_{\text{pl}})$ and $M \sim {\cal O}(M_{\odot})$.

In the region $\chi \ll 1$, the correction from the Schwarzschild solution can be 
calculated perturbatively by assuming a $\frac{1}{r}$ series expansion of $f,h$ 
and $\chi$. For example, the next to leading order correction can be obtained as
\begin{align}
 & f = 1 - \frac{2 GM}{r} + \frac{512 G^3 M^3}{M_L^6 r^9} +{\cal O}(r^{-10}),\label{fSch}\\
& h^{-1} = 1 - \frac{2 GM}{r} + \frac{2304 G^3 M^3}{M_L^6 r^9} + {\cal O}(r^{-10}),\label{hSch} \\
& \chi = \frac{48 G^2 M^2}{M_L^4 r^6} 
+  \frac{608256 G^4 M^4}{M_L^{10} r^{14}} +{\cal O}(r^{-15}).\label{chiSch}
\end{align}
Since the perturbative approach is only valid for $\chi \ll 1$, it is difficult to solve the 
equations of motion beyond $\chi \sim 1$ analytically. We will solve them 
numerically by using \eqref{fSch} - \eqref{chiSch} as the boundary conditions at some $r \gg r_L$. 

In order to see the effects of our modification, let us consider the case with 
$r_g \sim r_L$, which corresponds to an asymptotically Schwarzschild solution 
with a very small mass. As we will see below,  the behavior of the solution for 
$r_g < r_L$ is different from that for $r_L < r_g$. Let us investigate each case separately. 

\subsubsection*{Model1: Numerical solution with $r_L < r_g$}
\label{model1}

\begin{figure}[htbp]
\begin{center}
 \includegraphics[width=\hsize]{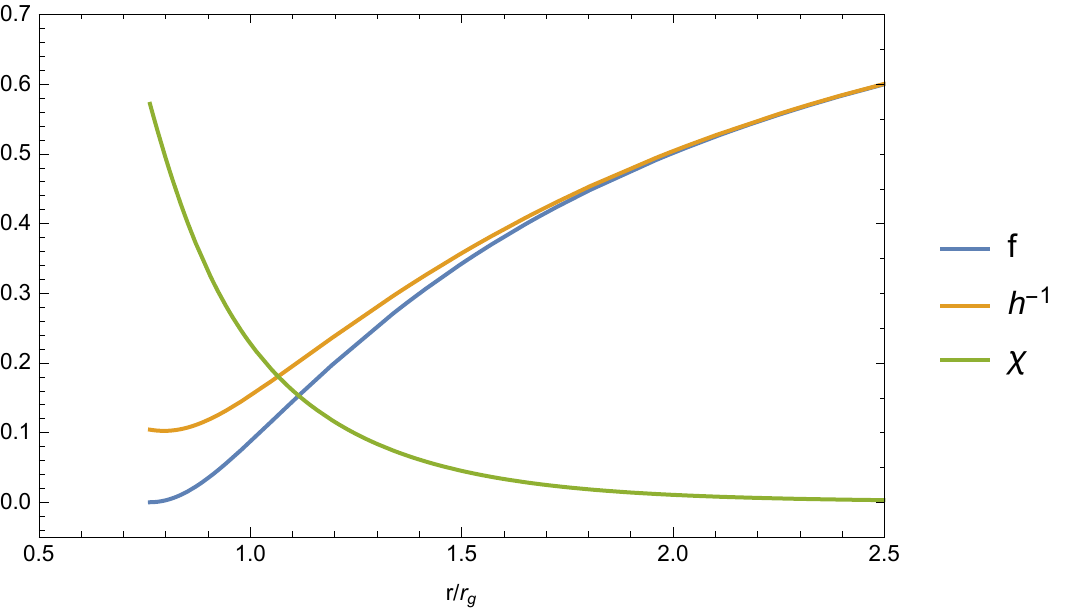}
\end{center}
\caption{Numerical solutions for Model 1}
\label{plot1}
\end{figure}

First let us focus on the case $r_L < r_g$, where higher derivative corrections become 
significant inside the event horizon expected from the asymptotic Schwarzschild space-time.
Concretely we set the parameter as $M_L = (G M)^{-1}$. For this parameter, $r_L$ 
becomes $r_L \sim 0.95 r_g < r_g$. The results of the numerical solution of the equations of motion
with this parameter choice are given by Fig. \ref{plot1}. In the numerical work,
we have used the initial conditions \eqref{fSch} - \eqref{chiSch} at $r=25 r_g $. 

\begin{figure}[htbp]
\begin{center}
 \includegraphics[width=\hsize]{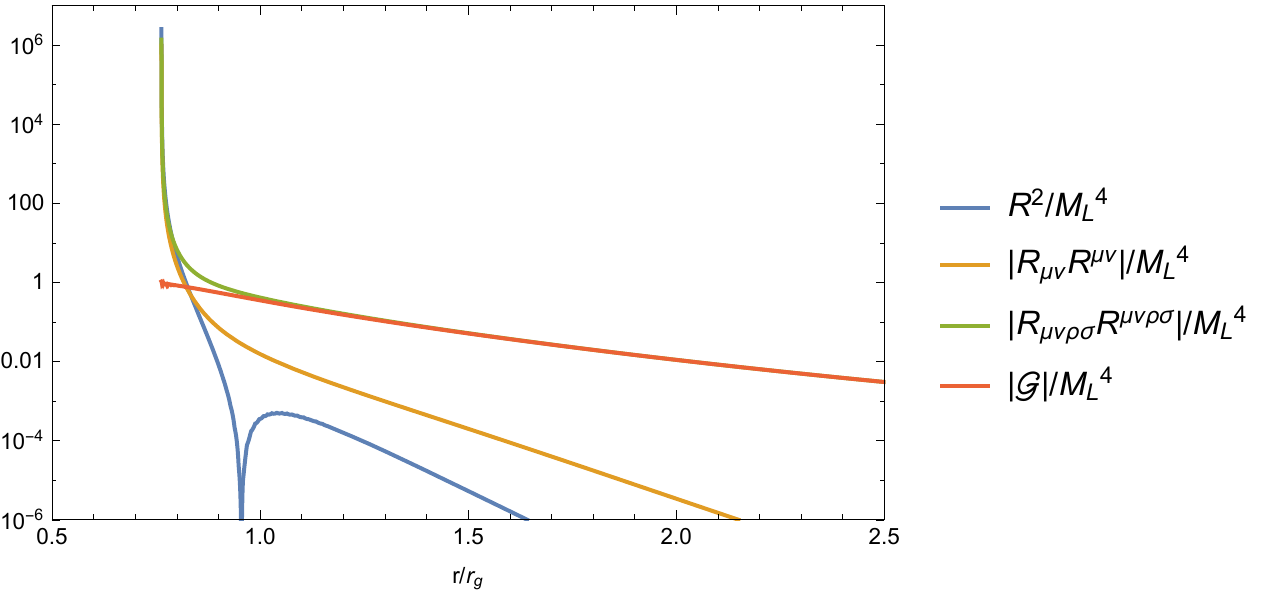}
\end{center}
\caption{Quadratic curvature scalars in Model 1}
\label{plot2}
\end{figure}

From the plot, we find that the numerical calculation stops at $r \sim 0.76 r_g$. At this point, $f$ vanishes  
but $h$ is finite. This point is the horizon. Its value has been shifted inwards by the
addition of higher curvature terms. More importantly, it has become a singular surface
in space-time. In order to clarify whether this point is a true singularity or an artificial singularity 
like a coordinate singularity, we plot the behavior of quadratic curvature scalars in Fig.~\ref{plot2}.  
From Fig.~\ref{plot2}, we find that both the curvature scalar $R, R_{\mu\nu}R^{\mu\nu}$ and $R_{\mu\nu\rho\sigma}R^{\mu\nu\rho\sigma}$ diverge at this point. Thus $r \sim 0.76 r_g$ is true 
curvature singularity. Note that although each quadratic curvature scalar is infinite, the 
Gauss-Bonnet term, which is the sum of these curvature scalars, is finite as expected.

The reason for the appearance of a singularity can be understood as follows. In Einstein gravity, 
the Schwarzschild solution written in terms of Schwarzschild coordinates has a coordinate 
singularity at the event horizon $r = r_g$, where $f$ vanishes and $h$ diverges while
maintaining the constraint $f h = 1$. The relation $f h=1$ ensures that $f = 0$ is not a 
physical singularity as can be seen by using Eddington-Finkelstein coordinates.  
Then once we include small correction terms in the gravitational action, the Schwarzschild 
solution is slightly modified. The important point is that, as one can see from 
Eqs.~\eqref{fSch} and \eqref{hSch}, the change in $f$ is generally different from that of 
$h^{-1}$, which leads to the breakdown of the relation $fh =1$ near the event horizon. 
Terms with $fh \neq 1$ lead to the event horizon of the original Schwarzschild space-time 
becoming a true curvature singularity as a consequence of the modification of the 
gravitational theory. This is the reason why our solution has a curvature singularity at a finite 
value of $r$. Since a similar singularity, called ``thunderbolt singularity'', was discussed 
in the context of the quantum effects in 1+1 dimensional space-time~\cite{Hawking:1992ti} 
and in Ho\v{r}ava-Lifshitz gravity~\cite{Misonoh:2015nwa}, we also call the singularity
we encounter here as a thunderbolt singularity. 

\subsubsection*{Model2: Numerical solution with $r_g < r_L$}
\label{model2}

\begin{figure}[htbp]
\begin{center}
 \includegraphics[width=\hsize]{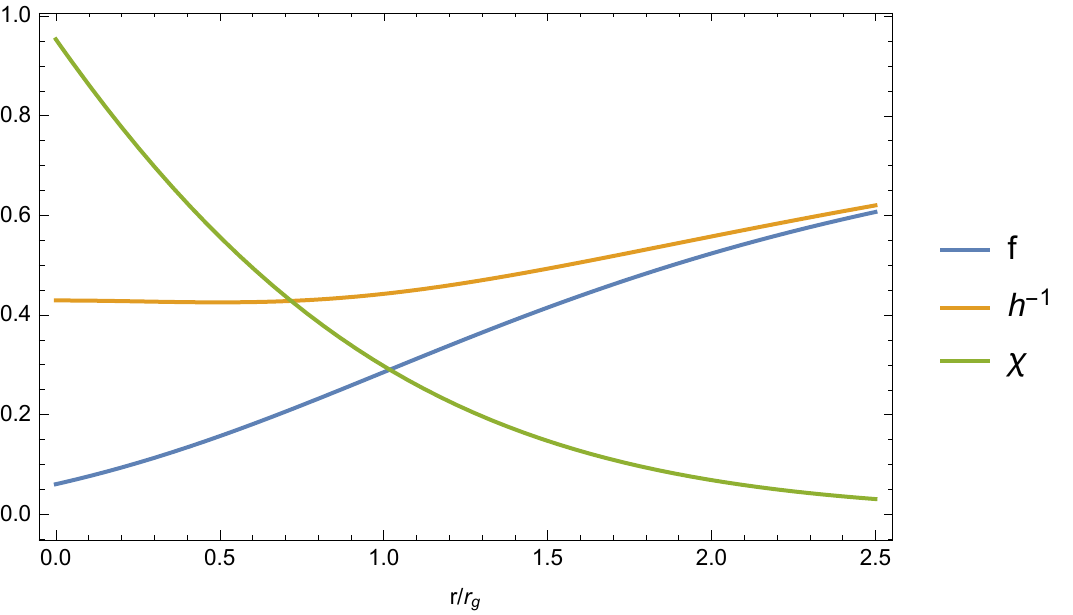}
\end{center}
\caption{Numerical solutions for Model 2}
\label{plot3}
\end{figure}

The thunderbolt singularity might not appear when $r_g < r_L$ because the effect of correction 
terms become significant at radii larger than where the event horizon  of the Einstein action
solution would be. Hence, it is possible that the horizon $f =0$ will not be reached (and hence
the singularity associated with this point would not be present).
To check our expectation, let us investigate the solution with the parameter choice
$GM = (2 M_L)^{-1}$, which corresponds to $r_L = 1.51 r_g > r_g$. The numerical solution 
is then given in Fig. \ref{plot3}.
Now we can continue the numerical calculation to $r = 0$ and both the horizon and the singularity 
at finite $r$ is successfully removed as expected.
Thus singularity coming from the breakdown of $fh=1$ is avoidable at least for 
asymptotically Schwarzschild space-time.

\begin{figure}[htbp]
\begin{center}
 \includegraphics[width=\hsize]{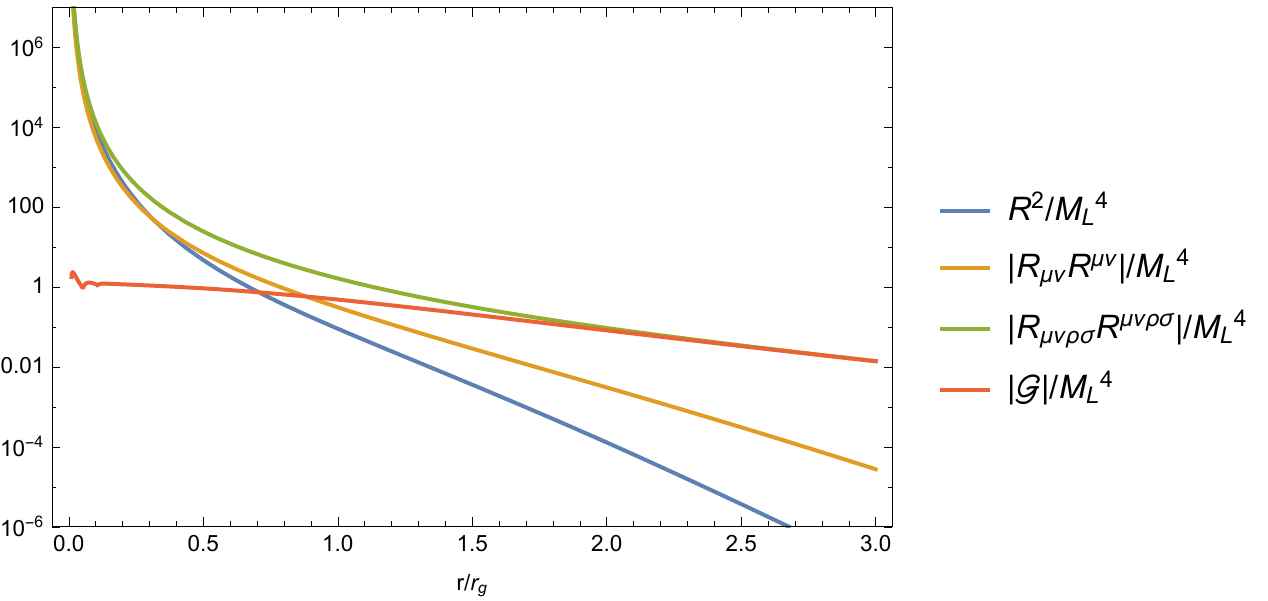}
\end{center}
\caption{Quadratic curvature scalars in Model 2}
\label{plot4}
\end{figure}

Then let us return our first question; Is the singularity at $r = 0$ is removed by limiting the 
Gauss-Bonnet term? The metric components are regular in the limit $r = 0
$, i.e. $f$ and $h$ are finite in this limit.
However as one can see from Fig.\ref{plot4}, the individual quadratic curvature
invariants which enter the Gauss-Bonnet term are infinite while the Gauss-Bonnet
term itself remains bounded.
Thus the original Schwarzschild singularity at $r=0$ still exists. In fact, it has become
a naked singularity since it is no longer shielded by a horizon.
The fact that the singularity at $r = 0$ is not removed should not be
too surprising because the requirement that ${\cal G}$ is finite is not sufficient to remove 
the divergence of other curvature scalars like $R, R_{\mu\nu}R^{\mu\nu}$. 
 
To summarize this section, we found that in a theory with bounded Gauss-Bonnet term 
there are two kinds of singularities which arise for spherically symmetric configurations, 
the thunderbolt singularity and the Schwarzschild singularity. The latter one could be 
removed by limiting other curvature scalars in addition to the Gauss-Bonnet term. 
However, we would have to go beyond the framework of known ghost-free higher 
derivative gravity models. Thus, to remove singularities with the limiting curvature 
mechanism would not be compatible with ghost-free requirement. In the following 
sections, we will discuss the singularity avoidance in wider class of theories setting
aside the issue of ghosts.

\section{Limiting both Ricci scalar and Gauss-Bonnet term}
\label{sec:GBR}

\subsection{How to ensure the finiteness of quadratic curvature invariants}
\label{sec:RGBLCH}

In the previous section, it was clarified that limiting only the Gauss-Bonnet term is 
not sufficient to remove the singularity at $r=0$.
Then, what is the condition to ensure finiteness of all quadratic curvature scalars at $r = 0$? 
Assuming the metric components are regular at $r =0$, they can be expanded 
in a Taylor series,
\begin{align}
 f = f_0 + f_1 r + f_2 r^2 + \cdots, \\
 h^{-1} = h_0 + h_1 r + h_2 r^2 + \cdots .
\end{align}
By plugging these expressions into Eq. \eqref{GB(r)}, 
the Gauss-Bonnet term is given by
\begin{align}
 {\cal G}(r) =& \frac{{\cal G}_0(f_0,f_1,f_2,h_0,h_1)}{r^2}\notag\\
& + \frac{{\cal G}_1(f_0,f_1,f_2,h_0,h_1,h_2)}{r} + {\cal O}(r^0),
\end{align}
where the expressions for ${\cal G}_0$ and ${\cal G}_1$ are 
\begin{align}
{\cal G}_0 =&
\frac{4 f_1 h_1}{f_0} + (h_0 -1) \frac{- 2  f_1^2 h_0  + 8 f_0 f_2 h_0  + 6 f_0 f_1 h_1}{f_0^2},\label{G0}\\
{\cal G}_1 =& - 2 \frac{f_1}{f_0} {\cal G}_0 +\frac{2 h_1 (f_1^2 + 8 f_0 f_2 + 3 f_0 f_1 h_1) + 8 f_0 f_1 h_2}{f_0^2} \notag\\
&+ (h_0 - 1) \frac{2 f_1^2 h_1 + 4 f_0(6 f_3 h_0 + 7 f_2 h_1 + 3 f_1 h_2)}{f_0^2}. \label{G1} 
\end{align}
The requirement that ${\cal G}$ is finite at $r = 0$ gives only 
two conditions for $f_m$ and $h_m$, namely ${\cal G}_0 = 0$ and ${\cal G}_1 = 0$, 
and these conditions are not sufficient to ensure that other curvature scalars are finite 
at $r =0$ . For example, The conditions ${\cal G}_0 = 0$ and ${\cal G}_1 = 0$ can be 
satisfied by appropriately choosing $f_0$ and $f_1$.
However, since the leading divergent term in the Ricci scalar, which is proportional to $r^{-2}$, 
comes from the first term in \eqref{R(r)}, it diverge unless $h_0 = 1$.
This is the reason why the divergence at $r=0$ appears in the framework of a $F({\cal G})$ theory.

Then let us now impose finiteness of $R$ in addition to that of ${\cal G}$. 
We can expand the Ricci scalar explicitly as
\begin{align}
 R(r) = - \frac{2(1- h_0)}{r^2} - \frac{2(f_1 h_0 + 2 f_0 h_1)}{f_0 r} + {\cal O}(r^0).\label{Rfh}
\end{align}
From the finiteness of $R$ at $r =0$ we obtain 
\begin{align}
 h_0 = 1,\qquad  h_1 = -  \frac{f_1}{2 f_0}.\label{finiteR}
\end{align}
Then, plugging these expression into Eq. \eqref{G0}, we get
\begin{align}
 {\cal G}_0 = - 2 \frac{f_1^2}{f_0^2} = 0.
\end{align}
Thus we find $f_1 = 0$. 
Moreover, from the expression \eqref{G1}, we can confirm that ${\cal G}_1$ also 
vanishes when the condition $f_1 = 0$, as well as \eqref{finiteR}, are satisfied. 
Without loss of generality, we can set $f_0 = 1$ by 
rescaling the time coordinate. 
Now the metric components are given by 
\begin{align}
 &f = 1 + f_2 r^2 + {\cal O}(r^3),\\
 &h^{-1} = 1 + h_2 r^2 + {\cal O}(r^3).
\end{align}
All scalar curvatures up to quadratic order are finite at $r = 0$, 
\begin{align}
 &R = -6 (f_2 + h_2) + {\cal O}(r), \\
 &\hat{R}_{\mu\nu}\hat{R}^{\mu\nu} = 3 (f_2 - h_2)^2 + {\cal O}(r), \\
 &{\cal G} = 24 f_2 h_2 + {\cal O}(r),
\end{align}
where $\hat{R}_{\mu\nu}$ is the trace-free part of the Ricci tensor defined 
by Eq.~\eqref{delRhat}. To summarize, if we impose the finiteness of $R$ and 
${\cal G}$, finiteness of all quadratic scalar curvatures at $r = 0$ is ensured as  long as the metric is regular at this point. 

\subsection{Spherically symmetric solutions with limiting $R$ and ${\cal G}$}
\subsubsection*{Model 3}
\label{model3}

In order to control both curvature scalars ${\cal G}$ and $R$,
we have to include $R$ as well as ${\cal G}$ in the argument of the arbitrary function $F$.
This is called an $F(R,{\cal G})$ theory,
\begin{align}
 {\cal L} = 
R + M_L^2 F\left(\frac{R}{M_L^2},\frac{{\cal G}}{M_L^4}\right)
.\label{FRG}
\end{align}
In Ref.~\cite{Yoshida:2017swb}, it was shown that non-singular cosmological solutions 
can be obtained in this framework. However $F(R, {\cal G})$ theory generally includes 
ghost degrees of freedom as can be explicitly seen by studying perturbations around 
Bianchi type I universes \cite{DeFelice:2010hg}. Here we pass over the ghost problem 
and focus only on the singularity problem. By comparing with \eqref{LCT}, the 
theory \eqref{FRG} can be regarded as a limiting curvature theory with $n=2$ and 
$I_1 = R/M_L^2, I_2 = {\cal G}/M_L^4$. Thus it can be written as 
\begin{align}
 {\cal L} =
R + M_L^2 \left( \chi_1 \frac{R}{M_L^2} + \chi_2 \frac{{\cal G}}{M_L^4} -  V(\chi_1,\chi_2) \right).\label{LRG}
\end{align}
For simplicity we focus only on the case of $V = V_1(\chi_1) + V_{2}(\chi_2)$. 
Variation with respect to $\chi_1$ and $\chi_2$ gives following equations to control 
$R $ and ${\cal G}$,
\begin{align}
 \frac{R}{M_{L}^2} = V'_1 (\chi_1), \qquad \frac{{\cal G}}{M_L^4} = V'_2(\chi_2).
\end{align}
Thus if we use potentials whose derivatives are finite for any value of $\chi_1$ and $\chi_2$, 
the theory only has solutions with finite values of $R$ and ${\cal G}$.

Since we have not solved the problem which arises at a horizon if the condition 
$fh = 1$ is violated, the thunderbolt singularity could still exist. First, however,
we shall focus only on the inside of the expected event horizon in order to see 
whether our mechanism to remove the Schwarzschild singularity works well or not. 
Thus we start the numerical calculations with the Schwarzschild boundary conditions 
at some value $r < r_g$.  We use the potentials
\begin{align}
 V_i(\chi_i) = \chi_i \arctan(\chi_i) - \frac{1}{2}\log(1 + \chi_i^2),
\end{align}
where
\begin{align}
V_i{}' = \arctan(\chi_i) .
\end{align}
Thus $V'$ is finite for any values of the $\chi_{i}$ fields. Then the analysis in the previous 
subsection implies that if $\chi_i \pm \rightarrow \infty$ at $r \rightarrow 0$ and if $f$ 
and $h^{-1}$ are regular there, the Schwarzschild singularity is removed. However, it
is non-trivial to show that this limit will be reached. Other singularities could appear
for finite values of the $\chi$ fields.
 
\begin{figure}[htbp]
\begin{center}
 \includegraphics[width=\hsize]{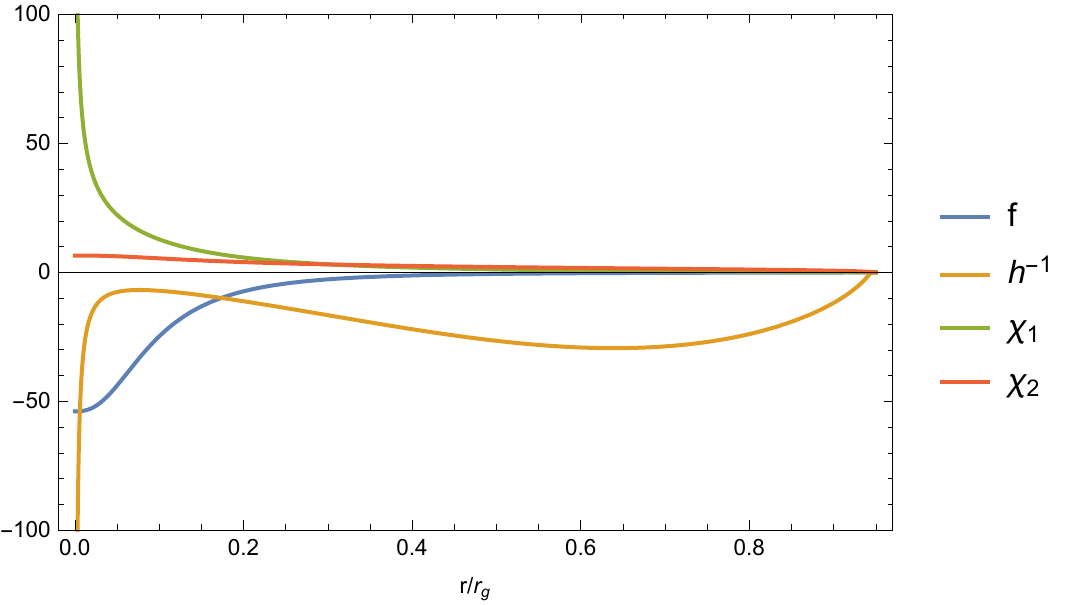}
\end{center}
\caption{Numerical solutions for Model 3}
\label{plotfrg1}
\end{figure}

In a similar way to what was done in subsection~\ref{Sec:SSAF}, we can derive the 
equations of motion by plugging the spherically symmetric ansatz for $f, h, \chi_1$ 
and $\chi_2$ into the action \eqref{LRG} and taking the variations with respect to each 
variable. Then the asymptotic Schwarzschild solution with mass $M$ is given by 
Eqs. \eqref{fSch}, \eqref{hSch}, \eqref{chiSch} for $\chi_2$ and we find
\begin{align}
 \chi_1 = \frac{2304 G^4 M^4}{M_L^8 r^{12}} +{\cal O}\left(\frac{1}{r^{13}}\right).
\end{align}
By using these solutions as our boundary conditions, we can numerically solve
the equations of motion, working from outside in (i.e. evolving the equations
towards smaller values of $r$). Fig.~\ref{plotfrg1} shows the numerical solutions 
for the parameter choice $ GM = 10 M_L^{-1}$ and starting with Schwarzschild 
boundary conditions at $r = 0.95 r_g$.
For this parameter choice, $r_L$ is given as $r_L = 0.21 r_g$. One can see that $h^{-1}$ 
diverges in the limit $r \rightarrow 0$. Thus, one of the assumptions made in 
section \ref{sec:RGBLCH}, which is that the metric components are regular at $r = 0$, is 
not satisfied.  Therefore, the question of whether the quadratic curvature scalars 
remain finite is still nontrivial in this setting. However, from the numerical results
we can compute these scalars. Fig.\ref{plotfrg2} presents the results for
the quadratic curvature scalars:
\begin{figure}
\begin{center}
 \includegraphics[width=\hsize]{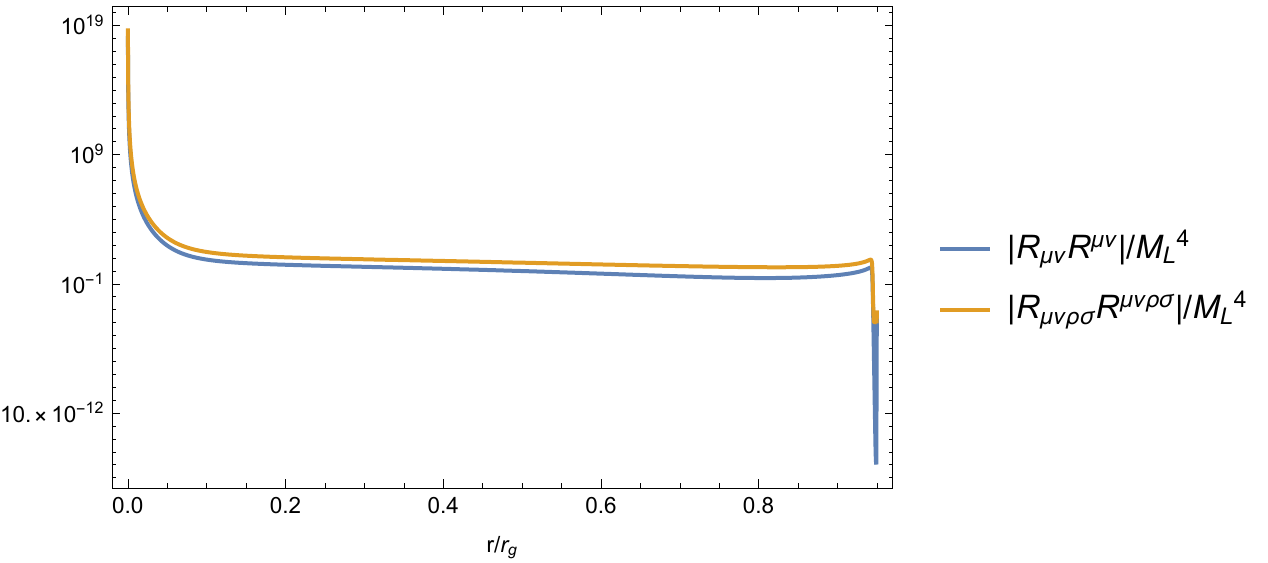}
\end{center}
\caption{Quadratic curvature scalars in Model 3}
\label{plotfrg2}
\end{figure}
We found that $R_{\mu\nu}R^{\mu\nu}$ diverges at $r = 0$. Therefore $ r = 0$ is 
still a singularity and we conclude that the Schwarzschild singularity cannot be 
removed even if we bound $R$ in addition to ${\cal G}$.

\section{Gravitational Theory with limiting Riemann Tensor}
\label{sec:limRiem}

\subsection{How to obtain $fh =1$}
\label{sec:fh=1}

We have seen that there are two kinds of singularities which come, respectively, from 
the lack of limiting curvature on one hand, and the violation of the condition $fh = 1$
on the other (recall that the latter condition was crucial is showing that
the horizon remains non-singular). In order to remove both singularities, we focus 
on theories that satisfy the following two conditions:
\begin{itemize}
 \item The theory has a sufficient number of bounded curvature invariants to ensure the 
 finiteness of all scalar curvatures up to quadratic order, namely $R,R_{\mu\nu}R^{\mu\nu},R_{\mu\nu\rho\sigma}R^{\mu\nu\rho\sigma}$ at $r = 0$ in
 order to remove the Schwarzschild singularity.
\item The theory admits only solutions which satisfy $f h = 1$. In this way, there is
a chance to avoid the thunderbolt singularity.
\end{itemize}
The first requirement would be satisfied if we control all components of the 
Riemann tensor. This is realized if we consider the theory with limited Riemann 
tensor given by \eqref{LCT}. 

Then let us investigate how the second condition can be realized in a theory 
with limited Riemann tensor.  We use the following spherically symmetric 
ansatz for $\chi_{\mu\nu\rho\sigma}$,
\begin{subequations} \label{sphericallysymchi}
\begin{align}
& \chi_{abcd} = A(r) f(r) h(r) \delta_{a[c} \delta_{d]b}, \\
& \chi_{tItJ} =  B_{tt}(r) f(r) r^2 \Omega_{IJ},  \\
&\chi_{rIrJ} =  B_{rr}(r) h(r) r^2 \Omega_{IJ},  \\
& \chi_{IJKL} = C(r) r^4 \Omega_{I[K}\Omega_{L]J},
\end{align}
\end{subequations}
which is compatible with the form of the Riemann tensor derived from our 
spherically symmetric metric \eqref{sphericallysymg}. Here the indices 
$I$ and $J$ run over $\theta$ and $\phi$, and the indices $a,b,c,d$ run over 
$t$ and $r$. For later convenience we introduce the following variables instead of 
$A,B_{tt},B_{rr}$ and $C$
\begin{align}
 &\chi = -A -4 (B_{tt} - B_{rr}) +C,\label{chi}\\
 &\xi = A - 2(B_{tt} - B_{rr}) -C,\label{xi}\\
 &\zeta = A + C,\label{zeta}\\
 &B = \frac{1}{2} (B_{tt} + B_{rr}),\label{LB}
\end{align}
where $\chi$ is the trace of $\chi_{\mu\nu\rho\sigma}$ as defined by \eqref{chi4=}. 
Now the components of $\hat{\chi}_{\mu\nu\rho\sigma}$ are functions of $\xi$, and 
the components of $\hat{\chi}_{\mu\nu}$ are functions of $\zeta$ and $B$.

The equations of motion can be derived by plugging the spherically symmetric 
ansatz \eqref{sphericallysymg} and \eqref{sphericallysymchi} into our action \eqref{LCT} 
and varying it with respect to $f,h,\chi,\xi,\zeta$ and $B$.
Since we defined $A,B_{tt},B_{rr}$ and $C$ so that the scalar quantities constructed 
from $\chi_{\mu\nu\rho\sigma}$ are independent of $f$ and $h$, the potential 
$V$ can be written as a function of $A, B_{tt}, B_{rr}$ and $C$, or as a function of 
$\chi, \xi, \zeta$ and $B$.  An important equation comes from the $B$ variation,
\begin{align}
 \frac{2}{ M_{pl}^2\sqrt{-g}}\left(\frac{\delta S}{\delta B}\right) = & \frac{4}{r h} \left( \log (fh)\right)' - V_{B} = 0.\label{eomBrrtt}
\end{align}
Then, if $V$ does not depends on $B$, i.e.  $V$ is a function like
\begin{align}
 V = V(\chi, \xi, \zeta),\label{assumptionV}
\end{align}
the solution of the equations of motion automatically satisfies
\begin{align}
 f h = 1.
\end{align}
Here we fixed the ambiguity of the integration constant by redefining the time coordinate $t$.
Thus, both challenges of preventing the divergence of quadratic curvature scalars at $r = 0$, 
and of removing the thunderbolt singularity which arises when $fh \neq 1$, are avoidable 
in the theory with limited Riemann tensor \eqref{LCT} with the potential \eqref{assumptionV}.  
However, this does not guarantee that no other singularities emerge. To study this question
we have to study the equations of motion in more detail.

\subsection{Asymptotically Schwarzschild solution}

Let us solve the equations of motion in the asymptotic region $ r \rightarrow \infty$.
The remaining equations of motion are given by
\begin{align}
&  A''+A' \left(\frac{4}{r}+\frac{ f'}{2 f}\right)
+\frac{ f' ( A-2 B_{rr}-2 B_{tt}-1)}{f r}
-\frac{4 B_{tt}'}{r} \notag\\
& \qquad +\frac{2 A -4 B_{tt} +  (C+1)(f^{-1} -1 )}{r^2}  - \frac{1}{2 f} M_L^2 V  = 0,\label{eomf} \\
&   \frac{A'}{2} f f' +\frac{ A-2 B_{rr}-2 B_{tt}-1}{r} f f' -\frac{4f^2  B_{rr}'}{r} \notag\\
&\ \ +\frac{ \left(-4 f^2  B_{rr}+  C
   \left(1+f^2\right)+ f(1-f)\right)}{r^2} - \frac{f}{2} M_L^2 V
 = 0,\label{eomh} \\
& \frac{2 (1-f) - 4 r f' - r^2 f''}{6 r^2} - M_L^2  V_{\chi} \notag \\
& \qquad \qquad \qquad 
= \frac{1}{6}R - M_L^2 V_{,\chi} = 0,\label{eomchi} \\
& \frac{-2(1-f) -2 r f' + r^2 f''}{3 r^2} - M_L^2 V_{,\xi}\notag\\
& \qquad \qquad \qquad  = 2C_{trtr} - M_L^2 V_{,\xi} =0,\\
& \frac{2(1 -  f) + r^2 f''}{r^2} - M_L^2 V_{,\zeta} \notag\\
&\qquad  \qquad \qquad = - 4 \hat{R}_{tt} - M_L^2 V_{,\zeta}= 0,\label{eomzeta}
\end{align}
where $A,B_{tt},B_{rr}$ and $C$ in Eqs. \eqref{eomf} and \eqref{eomh} are 
regarded as functions of $\chi, \xi, \zeta$ and $B$ is defined through \eqref{chi} - \eqref{LB}.

For simplicity, let us focus on the following form of the potential,
\begin{align}
 V(\chi,\xi,\zeta) = V_1(\chi) + V_2(\xi) + V_3(\zeta).
\end{align}
Then, from the expressions \eqref{eomchi} - \eqref{eomzeta}, one can see that the 
fields $\chi, \xi$ and $\zeta$ control $R, C_{trtr}$ and $\hat{R}_{tt}$, respectively. 

Assuming that the potentials have the following form for $\chi,\xi,\zeta \ll 1$,
\begin{align}
 &V_1(\chi) = \frac{1}{2}\chi^2 + a_4 \chi^4 + \cdots,\\
 &V_2(\xi) = \frac{1}{2} \xi^2 + b_4 \xi^4 + \cdots,\\
 &V_3(\zeta) =  \frac{1}{2}\zeta^2  + c_4 \zeta^4 \cdots,
\end{align}
the asymptotic Schwarzschild solution can be obtained perturbatively as
\begin{align}
 &f = 1 - \frac{2GM}{r} + \frac{896 b4 G^3 M^3}{285 M_L^4 r^7} + \cdots,\label{ASch-fRiemf}\\
 &\chi = \frac{ 896 b_4 G^3 M^3}{57 M_L^6 r^9} + \cdots ,\\
 &\xi = - \frac{4 GM}{M_L^2 r^3} + \frac{17152 b_4 G^3 M^3}{95 M_L^6 r^9}+ \cdots ,\\
 &\zeta = - \frac{8064 b_4 G^3M^3}{95 M_L^6 r^9} + \cdots,\\
 &B = \frac{B_1}{r}\frac{1}{1-\frac{2GM}{r}} + \frac{224 b_4 G^3 M^3}{285 M_L^4 r^7} \cdots  ,\label{ASch-fRiemB}
\end{align}
where $GM$ and $B_1$ are arbitrary constants.

\subsection{Reduction to first order differential equations} \label{sec:Hamiltonian}

In order to solve the equations of motion numerically, let us reduce them to first order form.
We can do this making use of the Hamiltonian formalism. Our equations of motion can 
be derived from the Lagrangian $L = 2{\cal L}/M_{pl}^2 \sin\theta $  which is given by
\begin{align}
 L =
&\frac{1}{6}\mathrm{e}^{-\Delta} r^2 f'(\chi' - 2 \xi' -3\zeta') \notag\\
&-\frac{1}{3}\mathrm{e}^{-\Delta} r f'(6 + \chi+4\xi+3\zeta) \notag\\
&+ \frac{2}{3}\mathrm{e}^{-\Delta} r f \Delta' (6 + \chi+ \xi + 12 B ) \notag\\
&-\frac{1}{3}\mathrm{e}^{-\Delta} f ( 6+\chi-2\xi+3\zeta) \notag\\
&+\frac{1}{3}\mathrm{e}^{\Delta} \left(
6+\chi-2\xi+3\zeta - 3 M_L^2 r^2 V(\chi,\xi,\zeta)
\right), 
\end{align}
where we introduced $\Delta$ as
\begin{align}
 \Delta = \frac{1}{2} \log (f h).
\end{align}
We regard $\Delta$ as one of the independent variables instead of $h$ and now we
 have 6 dynamical variables $q^{I} =\{f, \Delta, \chi, \xi,\zeta,B\}$.

Let us consider the Hamiltonian (regarding $r$ as a time coordinate).  By defining 
conjugate momenta $p_{I} = \partial L / \partial q^{I}{}'$ as usual, we obtain 
the following two relations between the momenta and the first derivatives of the variables,
\begin{align}
 & p_{f} =  - \frac{1}{3} \mathrm{e}^{-\Delta} r \left(6+\chi+4\xi+3\zeta \right)\notag\\
&\qquad \qquad \qquad  - \frac{1}{6}\mathrm{e}^{- \Delta} r^2 \left(-\chi'+2\xi'+3 \zeta'\right)\\
 & p_{\chi} = \frac{r^2}{6}\mathrm{e}^{-\Delta} f' \, .
\end{align}
We also obtain four primary constraints,
\begin{align}
&C_{\Delta} =  p_{\Delta} - \frac{2}{3}\mathrm{e}^{-\Delta} f r \left(6+\chi + \xi + 12 B\right) \approx 0,\label{C1}\\
&C_{\xi} = p_{\xi} + 2 p_\chi \approx 0,\\
&C_{\zeta} = p_{\zeta} + 3 p_{\chi} \approx 0,\\
&C_{B }= p_B \approx 0.\label{C4}
\end{align}  
Thus, the total Hamiltonian of this system is given by
\begin{align}
 H &= \sum_{I} p_I q^{I}{'} - L + \sum_{J =\{\Delta,\xi,\zeta,B\}} \lambda^{J} C_{J} \\
&= \frac{1}{3} \mathrm{e}^{\Delta}
\left(
-6-\chi+2\xi-3\zeta + \frac{18 p_f p_\chi}{r^2}
\right) \notag \\ & \qquad 
+\frac{1}{3}\mathrm{e}^{- \Delta} f (6 + \chi -2 \xi+3\zeta) \notag\\ &\qquad
 +\frac{2 p_\chi (6 + \chi +4\xi+3\zeta)}{r}\notag \\ & \qquad
+ \mathrm{e}^{\Delta} M_L^2 r^2 V(\chi,\xi,\zeta) +  \sum_{J =\{\Delta,\xi,\zeta,B\}} \lambda^{J} C_{J},
\end{align}
where $\lambda^{J}$ are Lagrange multipliers with respect to the primary constraints $C_J$.

Now the equations of motions of this system are given by the Hamilton equations
\begin{align}
 q^{I}{}' = \{q^I, H\},\  p_{I}{}' = \{q_I, H\},
\end{align}
where the Poisson bracket is defined by
\begin{align}
 \{F,G\} = \sum_{I} \frac{\partial F}{\partial q^{I}}\frac{\partial G}{\partial p_{I}}-\frac{\partial F}{\partial p_{I}}\frac{\partial G}{\partial q^{I}}.
\end{align}
Then the $r$ derivative of a function of $q^{I}$ and $p_I$ can be written in
terms of Poisson brackets as
\begin{align}
 \frac{d}{d r} F(r, q^{I},p_I) = \partial_{r} F + \{ F, H\}.
\end{align}
Since there are primary constraints \eqref{C1} - \eqref{C4} in this system, the 
variables $q^I$ and $p_I$ are not all independent.

Next we have to check the consistency of the constraints with the Hamilton equations.
The $r$ derivatives of $C_{\Delta}$ and $C_B$ can be calculated as
\begin{align}
 &C_\Delta{}' = \frac{\partial C_{\Delta}}{\partial r} + \{C_{\Delta}, H\} \notag\\
&\qquad = - 8 \mathrm{e}^{-\Delta} f r (\lambda_B - \hat{\lambda}_B(q^I,p_I,\lambda_{\xi},\lambda_{\zeta})) ,\label{Cdeltadot}\\
 &C_B{}' = \frac{\partial C_{B}}{\partial r} + \{C_{B}, H\}  = 8 \mathrm{e}^{- \Delta} f r \lambda_{\Delta} ,\label{CBdot}
\end{align}
where $\hat{\lambda}_{B}$ is given by
\begin{align}
 \hat{\lambda}_B =& - \frac{1 }{4} \left( \lambda_{\zeta} + \lambda_{\xi} + \frac{30 + 20 \xi + 5 \chi+9 \zeta +24 B}{6 r} \right) \notag\\
& - \frac{\mathrm{e}^{\Delta}}{2 f r^2} \left( f p_{f} +(6 + 12 B + \xi + \chi) p_{\chi} \right) \notag\\
& + \frac{\mathrm{e}^{2 \Delta}}{f r^3}\left(-18 p_f p_{\chi} + r^2 (6+3\zeta-2\xi+\chi) - 3 M_L^2 r^4 V\right).
\end{align}
Thus the consistency equations for $C_{\Delta}$ and $C_{B}$ fix two Lagrange multipliers to be
\begin{align}
 \lambda_\Delta = 0, \qquad \lambda_{B} = \hat{\lambda}_B,\label{lambdaDB}
\end{align}
unless $f = 0$. Since the consistency equations for $C_{\xi}$ and $C_{\zeta}$ do not include 
multiplier fields, they give two secondary constraints,
\begin{align}
& C_{\xi}' \approx 0 \notag \\
& \Leftrightarrow \ C^{(2)}_{\xi} = p_{\chi}  +  \frac{1}{12} \mathrm{e}^{\Delta}M_L^2 r^3 \left(
2 V_{1}' + V_{2}'
\right) \approx 0 ,\label{Cxidot}\\
& C_{\zeta}'  \approx 0 \notag \\
& \Leftrightarrow\  C^{(2)}_{\zeta} = f + \frac{1}{2}\mathrm{e}^{2 \Delta} \left(
-2 + M_L^2 r^2 \left(V_{1}' - V_{2}' +V_{3}'\right)
\right) \approx 0.\label{Czetadot}
\end{align}
Note that $V_{1}'$ represents $ V_{1,\chi}$ and not the $r$ derivative of $V_{1}$. 
The consistency equations for these secondary constraints are given as
\begin{align}
\begin{pmatrix}
 C^{(2)}_{\xi}{}' \\
 C^{(2)}_{\zeta}{}'
\end{pmatrix}
\approx
\mathrm{e}^{\Delta} M_L^2 r^2 \left( \mathbf{M}
\begin{pmatrix}
\lambda_{\xi}\\
\lambda_{\zeta}	       
\end{pmatrix}
- \begin{pmatrix}
F_{\xi}(q^I,p_I)\\
F_{\zeta}(q^I,p_I)	       
\end{pmatrix}
\right)
 \approx 0,\label{consistencyC2}
\end{align}
where the functions $F_{\xi}$ and $F_{\zeta}$ are given by
\begin{align}
 F_{\xi} &= \frac{3 V_2' + 2V_3' + 4(6+\chi + 4 \xi + 3\zeta+ 3 r^{-1} \mathrm{e}^{\Delta}  p_f ) V_1'' }{r} \\
 F_{\zeta} &= \frac{4 V_3' + 6 (6+\chi + 4 \xi + 3\zeta+ 3 r^{-1} \mathrm{e}^{\Delta}p_f ) V_1'' }{r}
\end{align}
 and the matrix $\mathbf{M}$ is given by 
\begin{align}
 \mathbf{M} =
\begin{pmatrix}
-4 V_1'' - V_2'' & - 6 V_1 ''  \\
 -6 V_1'' & - 9 V_1'' - V_3'' 
 \end{pmatrix}.
\end{align}
Thus, if the matrix $\mathbf{M}$ has an inverse, namely if its determinant is not zero,
\begin{align}
 \det \mathbf{M} =  9 V_1'' V_2''+ V_2''V_3'' + 4 V_3'' V_1'' \neq 0,\label{detM}
\end{align} 
then Eqs. \eqref{consistencyC2} can determine the remaining multipliers as
\begin{align}
 \begin{pmatrix}
  \lambda_{\xi} \\
  \lambda_{\zeta}
 \end{pmatrix}
 = \mathbf{M}^{-1} 
\begin{pmatrix}
 F_{\xi}\\
 F_{\zeta}
\end{pmatrix},\label{lambdaxz=}
\end{align}
and no more constraints appear.

Now we have 6 constraints $C_{\Delta}, C_{\xi}, C_{\zeta}, C_{B}, C_{\xi}^{(2)}$ and 
$C_{\zeta}^{(2)}$ which can be solved for $p_{\Delta},p_{\xi},p_{\zeta},p_B,p_{\chi}$ and $f$. 
Thus a complete set of equations of motion can be derived from the Hamilton equations 
for the remaining variables, $\Delta, p_f, \chi, \xi, \zeta$ and $B$, which now reduce to
\begin{align}
 &\Delta' = 0, \\
 &p_f{}' = - \frac{1}{3}\mathrm{e}^{-\Delta} (6 + \chi - 2\xi+3 \zeta), \\
 &\chi' = \frac{6 \mathrm{e}^{\Delta}p_f}{r^2}+\frac{2(6 +\chi+4\xi+ 3\zeta)}{r}+2 \lambda_{\xi} +3 \lambda_{\zeta},\\
 & \xi' = \lambda_{\xi}, \\
 & \zeta' = \lambda_{\zeta}, \\
 & B' = \lambda_{B},
\end{align}
where the Lagrange multipliers are determined from \eqref{lambdaDB} and \eqref{lambdaxz=}. 
Since $\Delta$ can be solved easily as $\Delta = 0$, i.e. $f h = 1$, we will solve the remaining 
5 equations numerically. Note that we have assumed $f \neq 0$ and $\det \mathbf{M} \neq 0$ 
when solving the equations \eqref{Cdeltadot},\eqref{CBdot} and \eqref{consistencyC2}. 
If either of these conditions is violated, the structure of the differential equations becomes 
singular in the sense that the number of independent initial conditions are changed. 
We can see this singularity as a divergence of the Lagrange multiplier $\lambda_I$ in the 
limits $f \rightarrow 0$ or 
$\det \mathbf{M} \rightarrow 0$.

\subsection{Numerical Calculation} \label{sec:solutionsfRiem}

\subsubsection*{Model 4}
\label{model4}

Now we are ready to study numerical solutions for given parameters and  potentials.
Let us consider the potentials,
\begin{align}
 V_i(x) = \frac{1}{2}\frac{x^2}{1+x^2}.
\end{align}
Since the derivatives of $V_i$,
\begin{align}
 V_i'(x) = \frac{x}{(1 + x^2)^2},
\end{align}
are finite for any $x$, solutions of the equations in this model have finite 
values of $R, \hat{R}_{tt}$ and $C_{trtr}$.
Since $V_i' \rightarrow 0$ in the limit where $\chi,\xi$ and $\zeta$ go infinity, 
solutions become non-singular Minkowski space-time in this limit. 

\begin{figure}[htbp]
\begin{center}
 \includegraphics[width=\hsize]{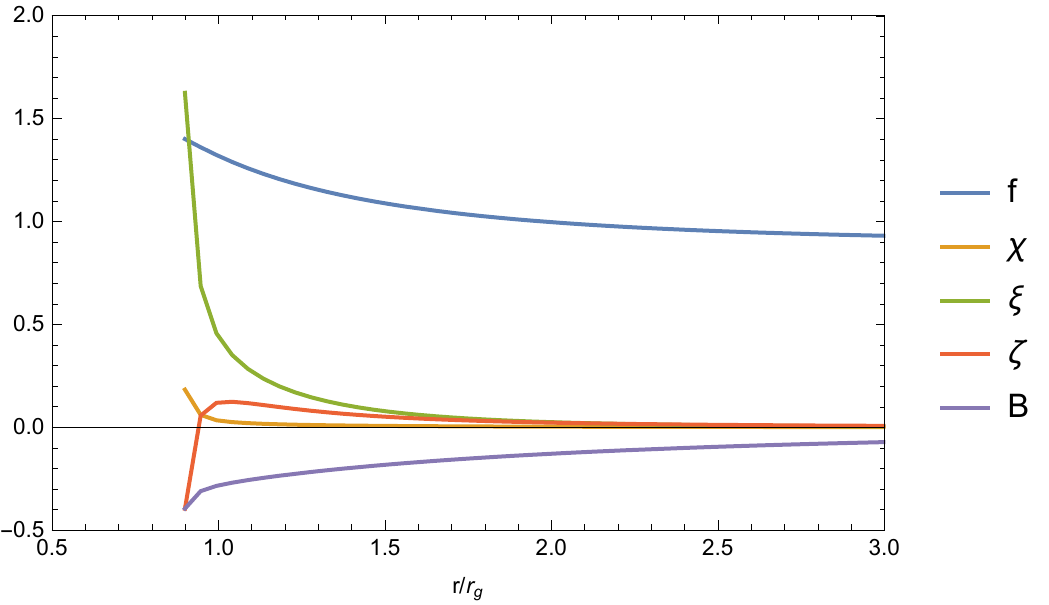}
\end{center}
\caption{Numerical solutions of Model 4}
\label{plot5}
\end{figure}

\begin{figure}[htbp]
\begin{center}
 \includegraphics[width=\hsize]{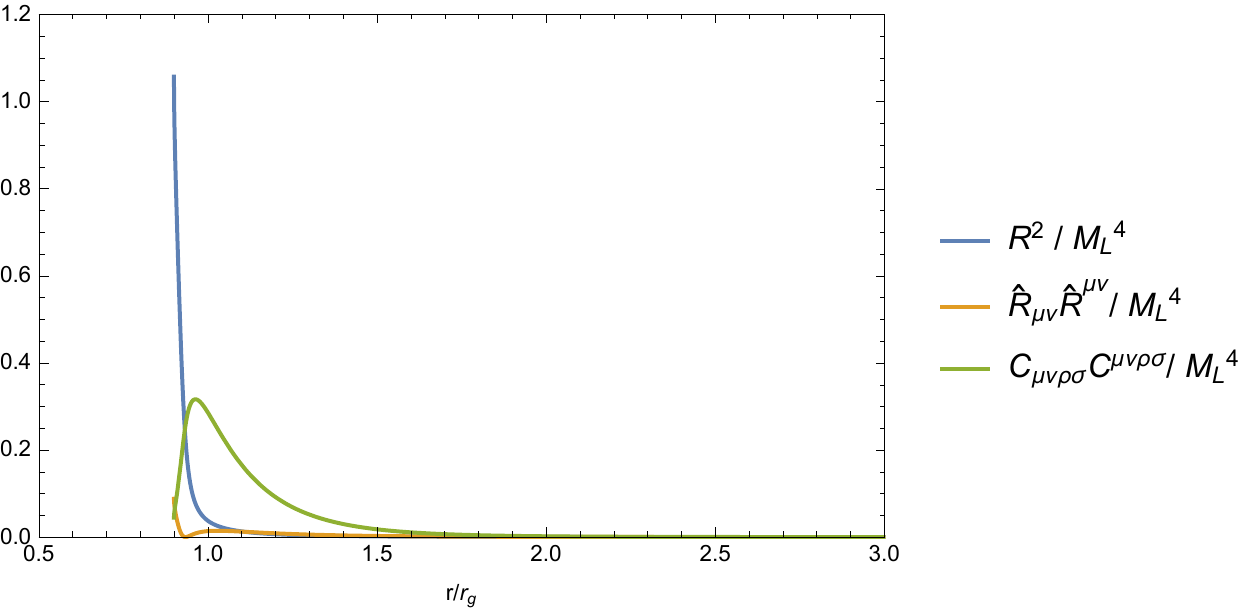}
\end{center}
\caption{Quadratic curvature scalars in Model 4}
\label{plot6}
\end{figure}

The numerical solution for the parameter choice $G M = M_{L}^{-1}$ 
(corresponding to $ r_L = 0.95 r_g$) and for Schwarzschild boundary conditions 
with $B_1 = 0$ at $r = 25 r_g$ is shown in Fig. \ref{plot5}. 
Even though all fields have finite values, a singularity appears at $r \sim 0.90 r_g$. 
There the curvature scalars $R$, $\hat{R}_{\mu\nu}\hat{R}^{\mu\nu}$ and $C_{\mu\nu\rho\sigma}C^{\mu\nu\rho\sigma}$ are finite as shown in fig. \ref{plot6}.
Then what is the origin of this singularity? 

\begin{figure}[htbp]
\begin{center}
 \includegraphics[width=\hsize]{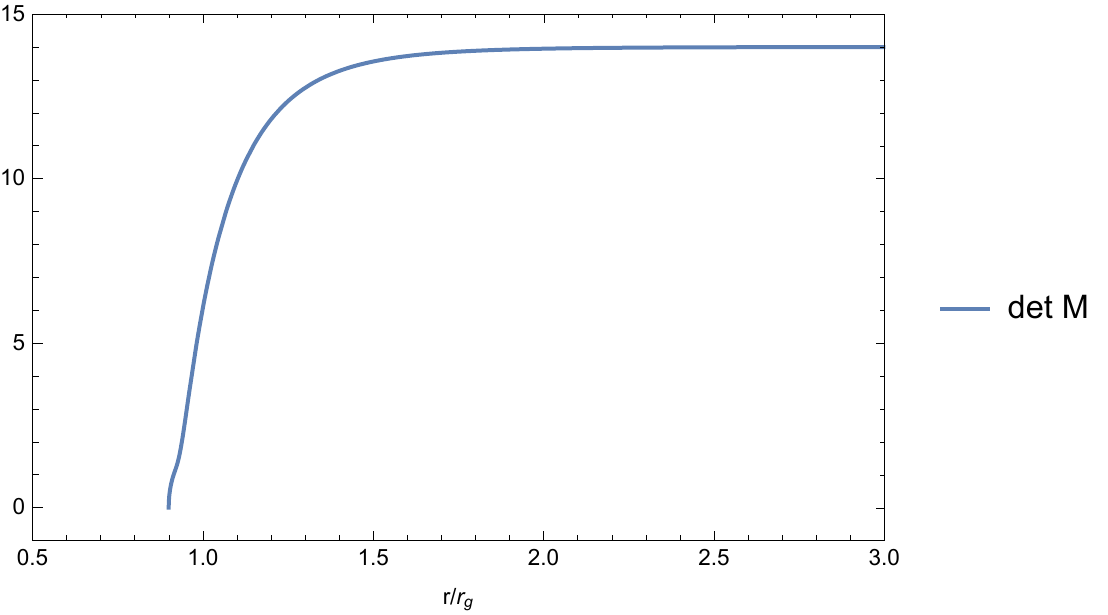}
\end{center}
\caption{Plot of det $\mathbf{M}$ in Model 4}
\label{plot8}
\end{figure}

The reason why we cannot extend our solution beyond $r \sim 0.90 r_g$ is because of 
the divergence of  $\chi', \xi', \zeta',$ and $B'$. Through the Hamilton equations, 
divergences of these quantities come from the divergences of Lagrange multipliers.
As mentioned, the Lagrange multipliers possibly become infinite when $f = 0$ or 
$\det \mathbf{M} =0$. Since $f \neq 0$ at the singularity, we conclude that the
singularity must be due to $\det \mathbf{M}$ vanishing at $r \sim 0.90 r_g$.  This 
is confirmed by the numerical plot of $\det \mathbf{M}$ given by Fig.\ref{plot8}.
Thus in this case, even though we can bound all quadratic curvature scalars, a singularity 
still appears because of the singular structure of the differential equationsin the limits 
$f \rightarrow 0$ or $\det \mathbf{M} \rightarrow 0$.
Roughly speaking $\chi, \xi$ and $\zeta$ represent curvature components through the 
equation \eqref{eomchi}- \eqref{eomzeta}. Thus the divergence of their derivative 
corresponds to that of curvatures (not of the curvature scalar, but to a
derivative thereof).

\begin{figure}[htbp]
\begin{center}
 \includegraphics[width=\hsize]{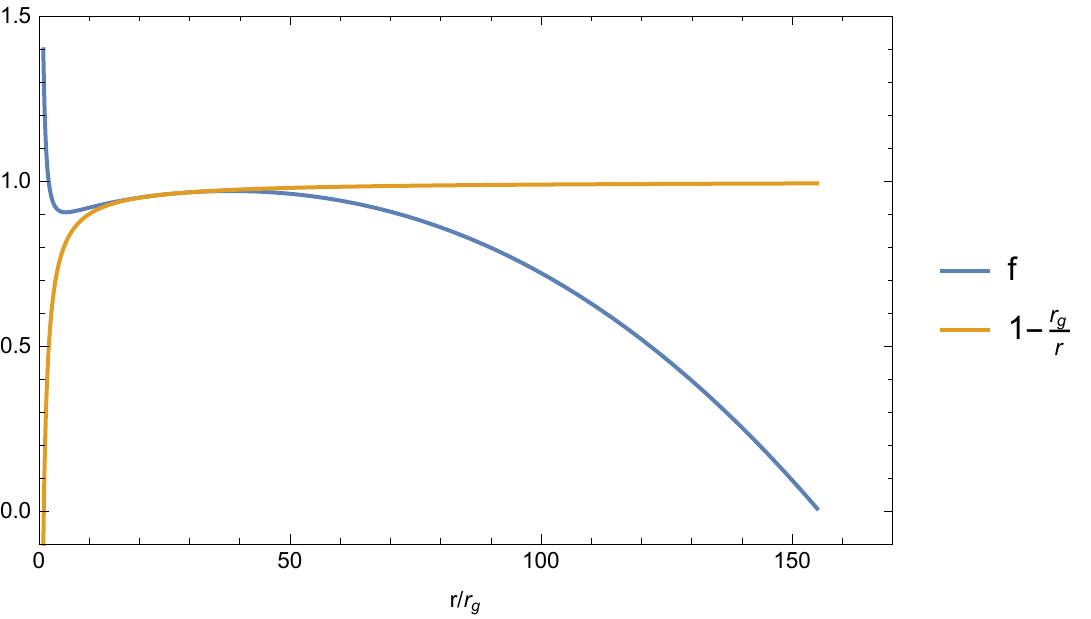}
\end{center}
\caption{Numerical instability of the asymptotic Schwarzschild solution}
\label{plot8-2}
\end{figure}

Note that the asymptotic Schwarzschild solution \eqref{ASch-fRiemf} - \eqref{ASch-fRiemB} 
is not a stable asymptote of the modified equations of motion. We can see this from fig.\ref{plot8-2}
where it is shown that if we integrate the equations in outward direction (towards larger
values of $r$), starting with Schwarzschild data at some finite $r$, that the solution then
runs away from the Schwarzschild solution. 
Thus our numerical solutions are not realistic even if there is no singularity
since they do not asymptote at large values of $r$ to an asymptotically
Minkowski space-time. In the current study, we pass over this stability 
problem as well as the ghost problem and focus only on the singularity problem.

The problems which we have encountered in this model may not be
general problems for this class of theories. Hence, it is useful to
study another model, a model in which the source of the singularity
in the previous model is cured.

\subsubsection*{Model 5}
\label{model5}

We will now numerically study solutions obtained for another potential. Since the 
singularity for Model 4 comes from a point in phase space where
$\det \mathbf{M} = 0$, it could be removed by considering a potential which 
enforces $\det \mathbf{M} \neq 0$.

Let us consider the following potentials,
\begin{align}
 V_i(x) = x \arctan(x) - \frac{1}{2}\log(1 + x^2) .\label{arctan}
\end{align}
Since $V_i''(x) =(1+x^2)^{-1}$ is positive for any finite $x$, $\det \mathbf{M}$ is also positive. 

\begin{figure}[htbp]
\begin{center}
 \includegraphics[width=\hsize]{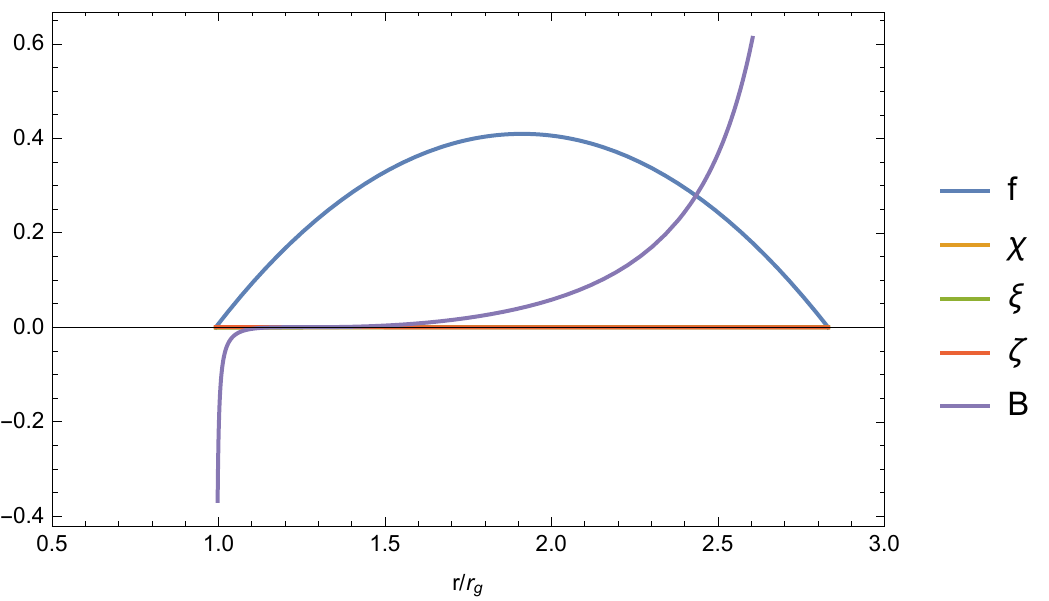}
\end{center}
\caption{Numerical solutions for Model 5}
\label{plot9}
\end{figure}

Here we make the parameter choice $GM = 20 M_L^{-1}$, which corresponds to $r_L = 0.13 r_g$.
Fig.\ref{plot9} presents the numerical solution for Schwarzschild boundary 
conditions at $r = 1.25 r_g$.
Now singularities appear at $r = 0.99 r_g$ and $r = 2.83 r_g$ where $f = 0$. 
Since the relation $fh = 1$ is satisfied by construction, 
$f = 0$ does not correspond
to a divergence of quadratic curvature scalars.

\begin{figure}[htbp]
\begin{center}
 \includegraphics[width=\hsize]{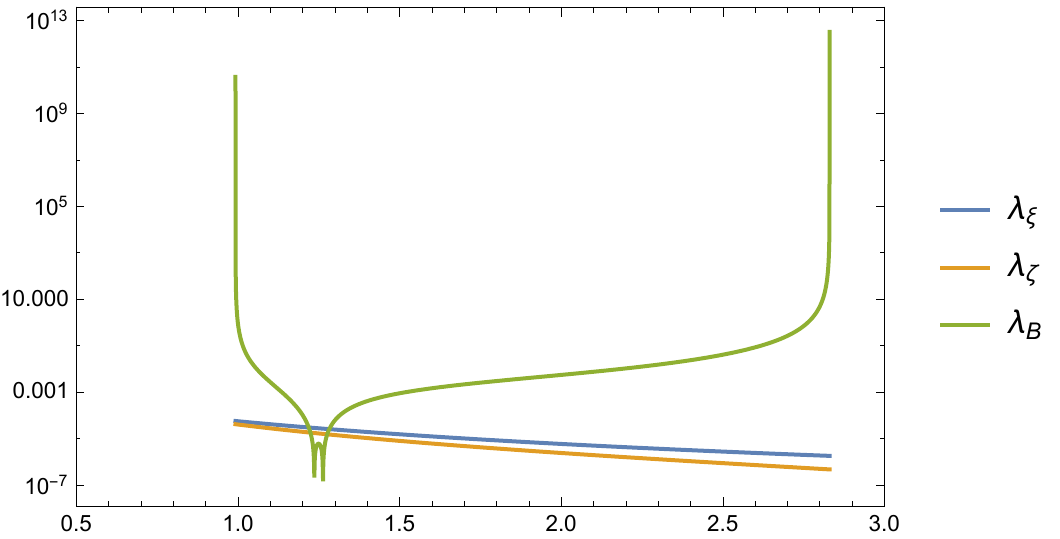}
\end{center}
\caption{Lagrange multipliers in Model 5}
\label{plot11}
\end{figure}

Nevertheless, $f = 0$ is still singular because it leads to
a singular structure of the differential equations
as what happens in the case $\det \mathbf{M} = 0$. One can see this from  
Fig.\ref{plot11}, where it is shown that $\lambda_B$ diverges at the points where $f = 0$.

\subsubsection*{Model 6}
\label{model6}

In the exact Schwarzschild space-time, $f$ vanishes at the event horizon $r = r_g$. Thus 
in order to avoid the appearance of $f = 0$, we need to have $r_g < r_L$ like in
Model 2 discussed in section \ref{model2}, though the required parameter choice is not 
natural for realistic situations.

\begin{figure}[htbp]
\begin{center}
 \includegraphics[width=\hsize]{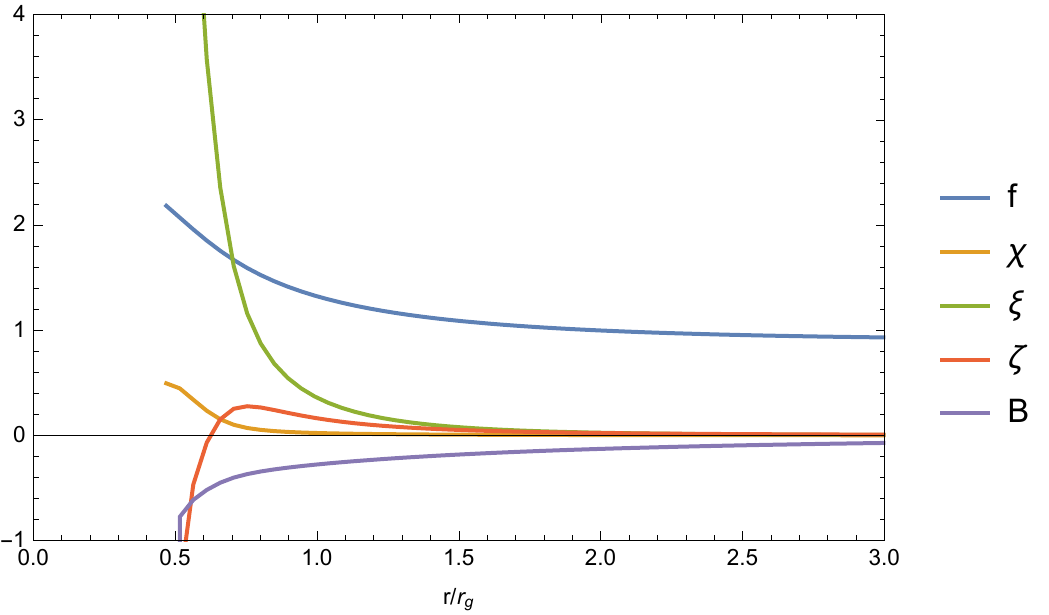}
\end{center}
\caption{Numerical solutions in Model 6}
\label{plot12}
\end{figure}

\begin{figure}
\begin{center}
 \includegraphics[width=\hsize]{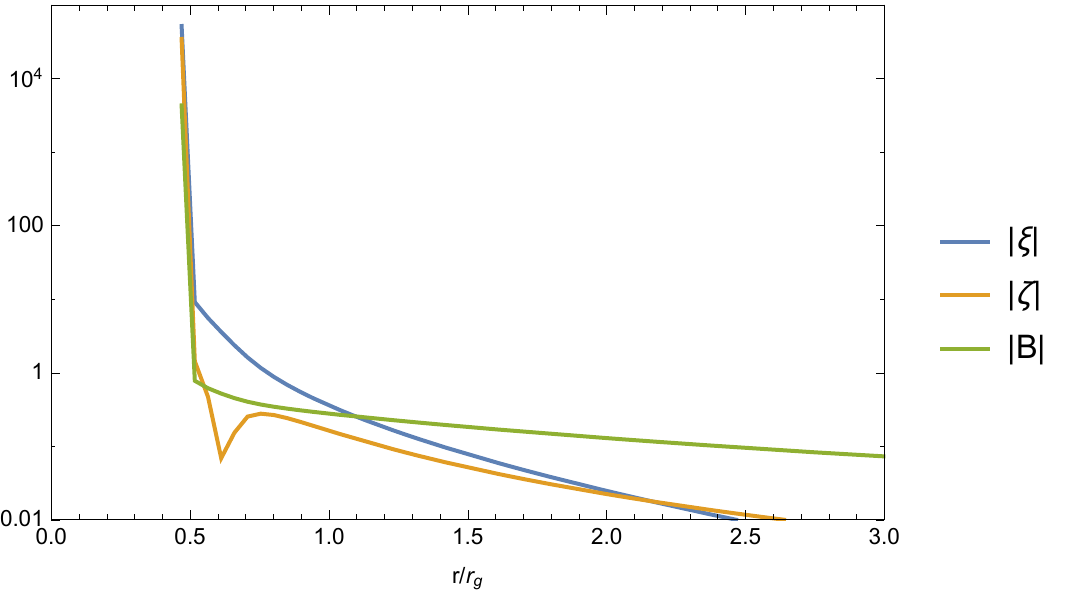}
\end{center}
\caption{Divergence of fields in Model 6}
\label{plot13}
\end{figure}

Let us investigate again a solution with the potential \eqref{arctan}. This time, let us 
make the parameter choice $GM = ML^{-1}$, which corresponds to $r_L = 0.95 r_g$. 
The results of the numerical solution with the Schwarzschild boundary conditions at 
$r = 25 r_g$ and with $B_1 = 0$ are shown in Figs.\ref{plot12} and \ref{plot13}. 
Here $f$ has finite values in the entire region but there is singularity at 
$r = 0.47 r_g$, where $\xi,\zeta$ and $B$ diverge.

\begin{figure}[htbp]
\begin{center}
 \includegraphics[width=\hsize]{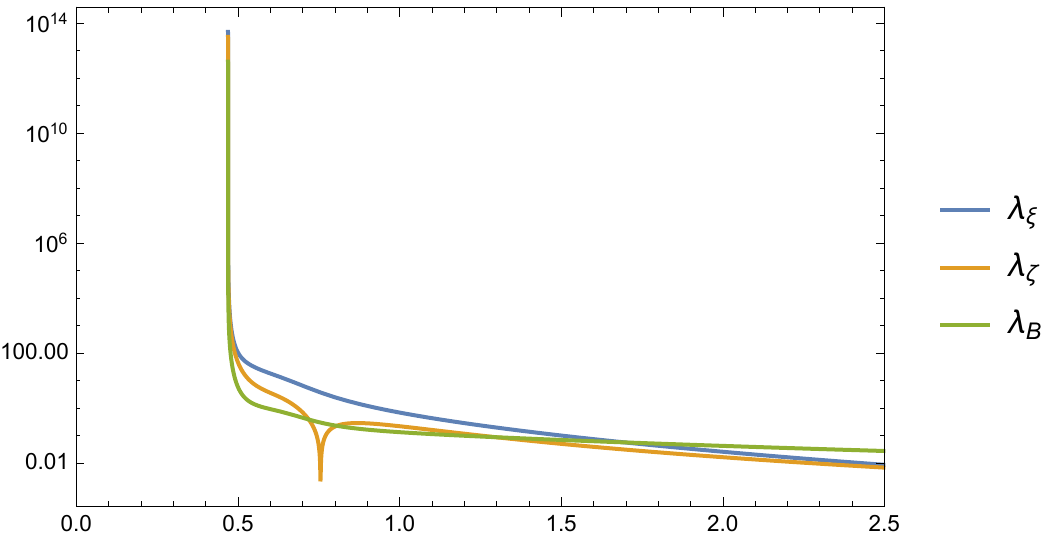}
\end{center}
\caption{Lagrange multipliers in Model 6}
\label{plot15}
\end{figure}

\begin{figure}
\begin{center}
 \includegraphics[width=\hsize]{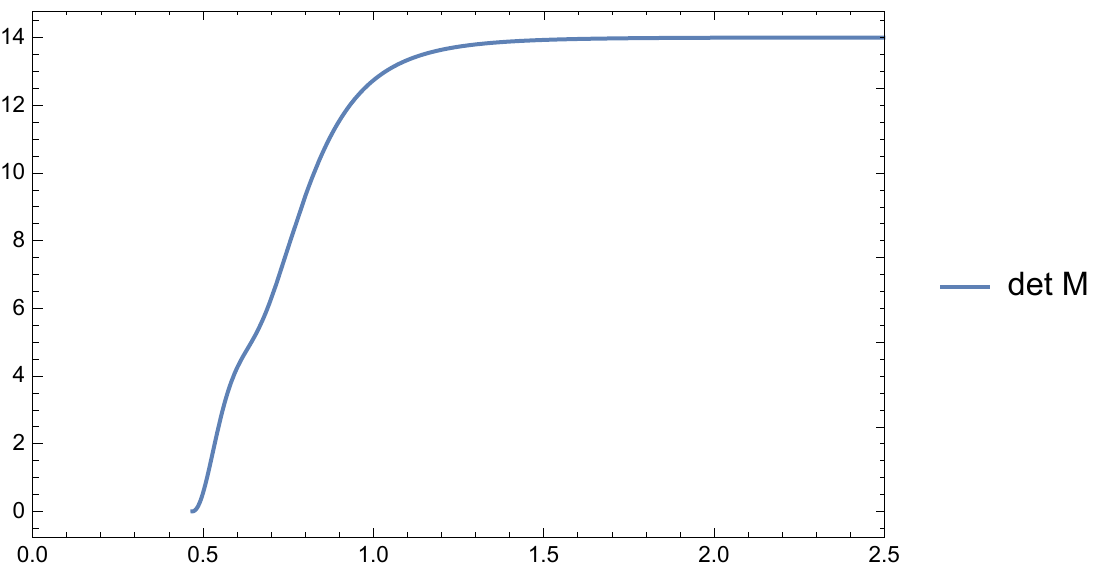}
\end{center}
\caption{Plot of $\det \mathbf{M}$ in Model 6}
\label{plot16}
\end{figure}

We used the potential \eqref{arctan} so that $det \neq 0$ for finite values of the
arguments, but $\det \mathbf{M}$ can vanish if the arguments ($\chi,\xi$ and $\zeta$) 
diverge. Actually, $\det \mathbf{M}$ vanishes and $\lambda$ diverges at the point 
$r = 0.47 r_g$ (See Figs. \ref{plot15} and \ref{plot16}).

One may think that the positivity of $\det \mathbf{M}$ would be ensured if we use a 
potential like $V''(x) > K$ with a positive constant $K$.
However such a potential cannot ensure the an overall upper bound on the curvature
invariants because $V'(x)$ is unbounded because 
$V'(x) > V'(x_0) + K (x - x_0) \rightarrow \infty$ in the limit $x \rightarrow \infty$. Thus, 
we see that limiting curvature invariants by our construction is not consistent with avoiding the 
{singular structure of the differential equations.

Note that since the metric component $f$ is well behaved at the singularity, the quadratic 
curvature scalars are finite. This is shown in Fig.\ref{plot14}.
\begin{figure}[htbp]
\begin{center}
 \includegraphics[width=\hsize]{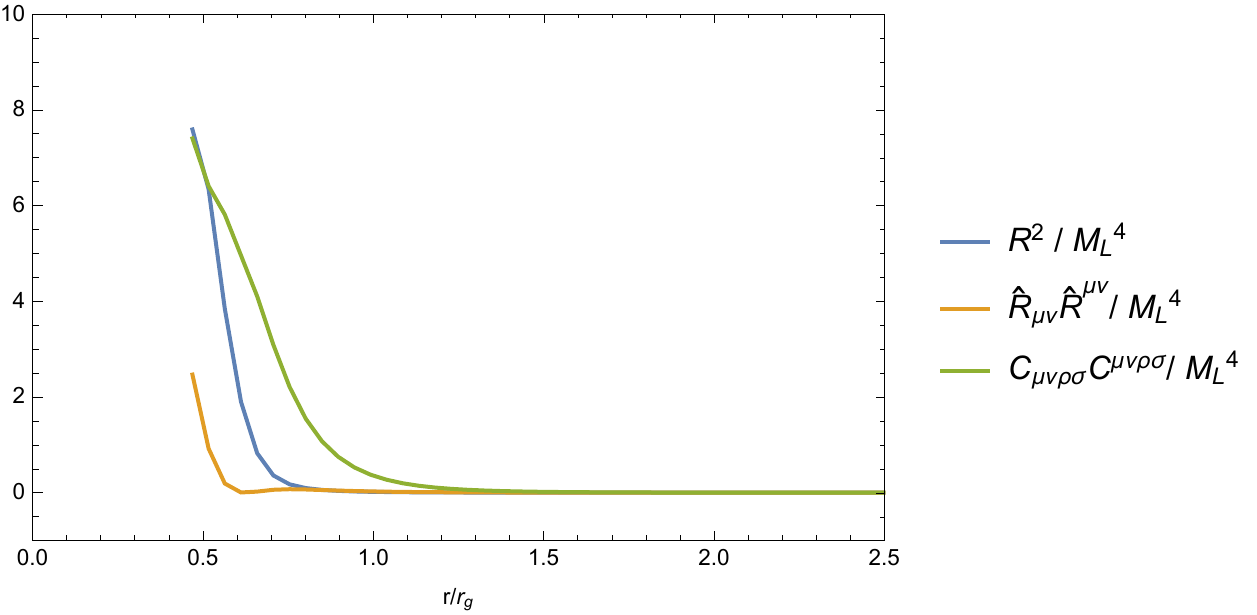}
\end{center}
\caption{Quadratic curvature scalars in Model 6}
\label{plot14}
\end{figure}

\section{Summary and Discussion}
\label{summary}

In this paper, we have discussed whether the Schwarzschild singularity can be
resolved in a theory with limited curvature invariants, a theory in which cosmological
singularities do not occur. In Section~\ref{ReviewofLC}, after reviewing the theory with 
bounded curvature scalars given by Eq.~\eqref{Slimcs}, the theory proposed in 
Refs.~\cite{Mukhanov:1991zn, Brandenberger:1993ef} which is able to produce non-singular 
cosmologies, we proposed a new theory in which all components of the curvature tensor 
are bounded by construction. The Lagrangian of this theory is given by Eq.~\eqref{LCT}. 
We also discussed the equivalence of these theories, \eqref{Slimcs} and \eqref{LCT}, with 
higher curvature metric theories, \eqref{F(I)} and \eqref{FRiem} respectively.  

In Section~\ref{sec:GB}, we investigated static, spherically symmetric solutions of the new
equations which reduce to Schwarzschild space-time at $r \rightarrow \infty$. First,
we considered Einstein gravity with bounded Gauss-Bonnet term given by~\eqref{SlimG}, 
which is a ghost free subclass of  limited curvature theories \eqref{Slimcs}.
We have given two numerical solutions (Models 1 and 2) for different parameter
choices, and found that there still exist singularities, in fact singularities of two kinds. 
One is the thunderbolt singularity found in Model 1 where the event horizon of the original 
Schwarzschild space-time becomes a curvature singularity. Some quadratic curvature 
invariants such as $R_{\mu\nu}R^{\mu\nu}$ diverge while the Gauss-Bonnet term is finite
since it is explicitly constrained by the construction. This singularity comes from the breakdown 
of the relation $f h = 1$, which holds in Einstein gravity. The other singularity found in Model 2 
is nothing but the original Schwarzschild singularity. The presence of the Schwarzschild 
singularity implies that limiting only the Gauss-Bonnet term is insufficient to remove the 
Schwarzschild singularity.  

Next, we investigated a theory in which both the Ricci scalar and the Gauss-Bonnet term 
are bounded by construction, a theory given by \eqref{LRG} (Section \ref{sec:GBR}).
However, the numerical solution of the equations of motion discussed in 
Section \ref{model3} (Model 3) shows that even in this framework the Schwarzschild singularity 
cannot be removed. 

Finally, we investigated a more general theory \eqref{LCT} in which all of the Riemann tensor
elements are bounded explicitly  (Section \ref{sec:limRiem}). 
In Section~\ref{sec:fh=1}, we found that the relation $f h = 1$ is automatically satisfied if we use 
the class of potentials given by Eq.~\eqref{assumptionV}. We derived the first order form of the 
equations of motion making use of the Hamiltonian formalism (Section \ref{sec:Hamiltonian})  
and found that the structure of the differential equations (e.g. the number of independent variables) 
is changed when either of the conditions $f \neq 0$ or $\det \mathbf{M} \neq 0$ is 
violated. 
We considered three types of specific models (Models 4, 5 and 6) (Section \ref{sec:solutionsfRiem}).  
All models yield some type of singularity. Model 4 leads to a singularity where $\det \mathbf{M}$ 
vanishes. Though all quadratic curvature invariants are finite at this singularity, as expected
from the construction, the additional degrees of freedom due to higher derivative interactions 
become strongly coupled at the singular point. In the case of the Models 5 and 6, we used 
a potential where $\det \mathbf{M} > 0$ for finite fields values. However, singularities remain 
in both models, again due to
the singular structure of the differential equations}. In Model 5, 
such a singularityappears
when $f = 0$, and in Model 6 it arises because $\det \mathbf{M}$ approaches
$0$ when the fields $\xi, \zeta$ and $B$ diverge. Thus the singularity at finite $r$ 
still remains even though the quadratic curvature invariants are finite at the
singular point. 

To summarize, we numerically studied the equations of motion for a
spherically symmetric ansatz for the fields in various theories in which the
curvature is bounded by construction. But in all cases, the solutions have singularities 
of various types. The results are summarized in Table \ref{table1}.
\begin{widetext}
\begin{center}
\begin{table}[htbp]
\begin{tabular}{l|cccl}
 Models & Theory & Position of Singularity & Quadratic Curvatures & Origin of Singularity\\
\hline
Model 1 (Sec. \ref{model1}) & $F({\cal G})$  & $f = 0$ & $R, R_{\mu\nu}R^{\mu\nu} \rightarrow \infty$ &  $fh \neq 1$\\
Model 2 (Sec. \ref{model2}) & $F({\cal G})$  & $r = 0$ & $R, R_{\mu\nu}R^{\mu\nu} \rightarrow \infty$ & lack of limited curvatures \\ 
Model 3 (Sec. \ref{model3}) & $F(R, {\cal G})$ & $r = 0$ & $R_{\mu\nu}R^{\mu\nu} \rightarrow \infty$ & lack of limited curvatures\\ 
Model 4 (Sec. \ref{model4})  & $F(R_{\mu\nu\rho\sigma})$& $\det \mathbf{M} = 0$& finite & singular structure of differential equations\\ 
Model 5 (Sec. \ref{model5}) & $F(R_{\mu\nu\rho\sigma})$& $f = 0$& finite & singular structure of differential equations\\ 
Model 6 (Sec. \ref{model6}) & $F(R_{\mu\nu\rho\sigma})$& $\xi,\zeta,B \rightarrow \infty (\det{\mathbf{M}} \rightarrow 0) $& finite & singular structure of differential equations
\end{tabular}
 \caption{Summary of Numerical Solutions}
\label{table1}
\end{table}
\end{center}
\end{widetext}
We would like to emphasize that our analysis in Section \ref{sec:limRiem} gives a concrete 
counter-example to the strong form of the ``limiting curvature hypothesis'' according to
which general singularities could be avoided by using a Lagrangian in which the
curvature is explicitly bounded by construction. Thus, the limiting curvature hypothesis 
does not resolve general singularities, and another principle is required if we want to
construct an effective theory of gravity in which no singularities arise.

In the models in Section \ref{sec:limRiem}, the origin of the singularity was the dynamics 
of additional degrees of freedom. Since limiting curvature theories are essentially 
higher derivative theories as shown in Section \ref{ReviewofLC}, it is difficult to 
sufficiently well constrain the dynamics of such additional degrees of freedom. 
One possible avenue would be making use of the Palatini or metric affine formalism~\cite{Olmo:2011uz}, 
where the connection which determines the curvature tensor is independent 
of the metric tensor. This is a promising avenue because higher curvature gravity in 
the Palatini formalism does not include additional ghost degrees of freedom~\cite{Afonso:2017bxr}.

It is fair to say that our models are toy models and there would be many problems even if 
the Schwarzschild singularity could have been removed. For example, we have 
not addressed the problem of ghosts, and the numerical stability of the equations
(the question of whether asymptotically flat solutions are stable in the large $r$ limit. 
The models in Sections \ref{sec:GBR} and \ref{sec:limRiem} in general have ghost 
degrees of freedom. One way to justify such a higher derivative gravity model would 
be to regard the theory as a low energy effective theory after some heavy fields have 
been integrated out.  Naively speaking, since ghost modes appear because of 
higher derivative interactions which are suppressed by $M_L$, the mass of the 
ghost modes should be of the order of $M_L$. Then we need to regard our theory as 
an effective field theory valid at energies $E \ll M_L$. Since the curvature scale can be 
controlled by hand in our framework, a self-consistent procedure would to bound 
the curvatures to values corresponding to an energy scale smaller than $M_L$ by 
choosing a suitable potential. In this way, the extra terms in our gravitational action
would be within the energy range of the effective field theory, while the ghost
degrees of freedom would not.

Though our analysis is not a ``no-go'' result for non-singular black hole solutions 
in an approach in which the curvature is bounded by construction, we  conclude that 
the singularities cannot be removed generally if only the curvature is limited to finite values. 

\begin{acknowledgments}
We would like to thank Jerome Quintin for discussions.  D.Y. is supported by the Japan Society for the Promotion of Science (JSPS) Postdoctoral Fellowships for Research Abroad. R.B. is supported in part by a Discovery Grant by the Canadian NSERC, and by funds from the Canada Research Chairs program. R.B. thanks Lavinia Heisenberg for discussions.
\end{acknowledgments}

\bibliography{ref}
\end{document}